%% file: lownu_minerva.tex
\newcommand{\DoPrePrint}{0} 

\documentclass[%
reprint,
superscriptaddress,
showpacs,
preprintnumbers,
showkeys,
nofootinbib,
 amsmath,amssymb,
 aps,
 prd,
floatfix,
]{revtex4-1}

\usepackage{xspace}
\usepackage{graphicx}
\usepackage{array}
\usepackage{dcolumn}
\usepackage{bm}
\usepackage{rotating}
\usepackage{multirow}

\usepackage{hyperref}

\usepackage[mathlines]{lineno}
\ifnum\DoPrePrint=1
\linenumbers\relax 
\fi

\newcommand\minerva  {MINERvA\xspace}
\newcommand\numu     {$\nu_{\mu}$\xspace}
\newcommand\numubar  {$\bar{\nu}_{\mu}$\xspace}

\newcommand\LowNu    {Low-$\nu$\xspace}
\newcommand\lownu    {low-$\nu$\xspace}

\newcommand\enu      {$E_\nu$\xspace}
\newcommand\signu    {$\sigma_\nu$\xspace}
\newcommand\GENIE  {\textsc{genie}\xspace}
\newcommand\GEANT  {\textsc{geant4}\xspace}
\newcommand\FLUKA  {\textsc{fluka}\xspace}

\begin{document}

\preprint{FERMILAB-PUB-16-368-ND}

\title{Measurements of the Inclusive Neutrino and Antineutrino Charged Current Cross Sections in MINERvA Using the Low-$\nu$ Flux Method} 

\input{authors}

\date{\today}

\begin{abstract}
The total cross sections are important ingredients for the current and future neutrino oscillation experiments. We present measurements of the total charged-current neutrino and antineutrino cross sections on scintillator (CH) in the NuMI low-energy beamline using an {\em in situ} prediction of the shape of the flux as a function of neutrino energy from 2--50\,GeV. This flux prediction takes advantage of the fact that neutrino and antineutrino interactions with low nuclear recoil energy ($\nu$) have a nearly constant cross section as a function of incident neutrino energy. This measurement is the lowest energy application of the \lownu flux technique, the first time it has been used in the NuMI antineutrino beam configuration, and demonstrates that the technique is applicable to future neutrino beams operating at multi-GeV energies.  The cross section measurements presented are the most precise measurements to date below 5~GeV.  

\end{abstract}

\maketitle

Current and future neutrino oscillation experiments that use neutrino energies from 1-10 GeV\,\cite{nova-tdr,NuMI-NIM,T2K-flux,DUNE-CDR} encounter a number of scattering processes that occur with comparable probabilities. Over the last decade, a series of experiments including K2K\,\cite{K2K-QE}, SciBooNE\,\cite{sciboone-inclusive}, MiniBooNE\,\cite{miniboone-quasi}, and T2K\,\cite{T2K-xsec-2,T2K-xsec-1} have explored neutrino cross sections in the sub-GeV to few GeV region and have largely focused on exclusive processes. However, since the NOvA\,\cite{nova-tdr} and DUNE\,\cite{DUNE-CDR} detectors are fully active, those experiments plan to use as much of the total neutrino interaction rate as possible. Therefore precise knowledge of the inclusive charged-current cross section is becoming increasingly important.  

Cross section measurements are challenged by imperfect knowledge of the incoming neutrino and antineutrino fluxes. Fluxes can be predicted using detailed beamline simulations benchmarked with an array of external hadron production data\,\cite{T2K-flux,NOMAD-xsec,Leo-thesis}, constrained using {\em in situ} tertiary muon production measurements\,\cite{Loiacono-thesis}, or measured using well understood processes such as neutrino-electron scattering\,\cite{minerva-nu-e-flux}. Because the flux is the largest uncertainty in most cross section measurements, constraining the flux using multiple independent techniques is important. 

The \lownu method for flux estimation was proposed by Mishra\,\cite{Mishra}, pioneered by the CCFR collaboration\,\cite{CCFR-xsec}, and used by the NuTeV\,\cite{NuTeV-xsec} and MINOS\,\cite{MINOS-xsec} collaborations.  The inclusive charged-current scattering cross section for neutrinos can be expressed in terms of neutrino energy (\enu), the energy transferred to the nucleus (recoil energy or $\nu$), and the Bjorken scaling variable ($x$) as
\begin{widetext}
\begin{equation}
    \frac{\mathrm{d}\sigma}{\mathrm{d}\nu} =
    \frac{G^2_F M}{\pi}
    \int_0^1 \bigg(
        F_2
        - \frac{\nu}{E_\nu}  \left[F_2 \mp xF_3\right]
        + \frac{\nu}{2E_\nu^2}  \left[\frac{Mx(1-R_L)}{1+R_L}F_2\right] 
        + \frac{\nu^2}{2E_\nu^2}  \left[\frac{F_2}{1+R_L} \mp xF_3\right]
    \bigg)\,\mathrm{d}x,
    \label{eq:dsigma-dnu}
\end{equation}
\end{widetext} 
where $G_F$ is the Fermi constant, $M$ is the struck nucleon's mass, $F_2$ and $xF_3$ are structure functions, $R_L$ is the structure function ratio $F_2/(2xF_1)$, and the ``$+$'' is for \numu and the ``$-$'' is for \numubar\,\cite{Josh-thesis}.  If one limits the final-state phase space to events with $\nu$ less than some cutoff $\nu_0$ (where $\nu_0 \ll E_\nu$) the terms proportional to $\nu/E_\nu$, $\nu/E_\nu^2$, and $\nu^2/E_\nu^2$ are small, yielding a cross section that is approximately constant as a function of neutrino energy. The \lownu restricted cross section deviates from constant both due to the finite value of $\nu_0$ in practical applications and due to the small $Q^2$ dependence (Bjorken scaling violation) of the structure functions\,\cite{Bodek:2012uu}. While the formulation in Eq.\,\ref{eq:dsigma-dnu} is only strictly true in the DIS regime, the \lownu technique is still applicable at lower momentum transfer since one is restricting the sample to a uniform final-state phase space and a correspondingly small range of $Q^2$. Cross sections for neutrino-nucleon scattering from the quasi-elastic (QE) regime through resonance production do have energy dependencies, especially for \enu = 1\,GeV. Corrections for these effects and the associated systematic uncertainties will be discussed later in this paper. 

The shape of the neutrino charged-current event spectrum for $\nu<\nu_0$ can therefore be used to determine the shape of the neutrino flux as a function of \enu, with the absolute normalization determined by scaling the flux such that the extracted inclusive cross section matches the world average at larger \enu where it is most precisely measured. This paper reports measurements of the shape of the neutrino and antineutrino charged-current inclusive cross sections using neutrino fluxes determined by the \lownu interactions with data from the MINERvA experiment\,\cite{Josh-thesis}. Enabled by the low threshold and good energy resolution of the detector, these measurements cover an energy range that extends to lower energies than previous measurements\,\cite{PDG}, and represent an important demonstration of the \lownu flux technique for future experiments operating in the few-GeV region. 

The MINERvA detector is a finely-segmented solid-scintillator detector consisting of a fully-active tracker region surrounded by electromagnetic and hadronic calorimeters\,\cite{MINERvA-NIM,DAQ-NIM}. The events for this analysis occur in the 5.57-ton fiducial region of the tracker. The tracker is composed, by weight, of 7.5\% hydrogen, 88\% carbon, 3.2\% oxygen, and smaller quantities of more massive nuclei, and has a 15\% excess of protons over neutrons. The magnetized MINOS near detector\,\cite{MINOS-NIM}, located 2~m downstream of the MINERvA detector, serves as a tracking muon spectrometer. 

The detectors are located in the NuMI neutrino beamline at Fermilab\,\cite{NuMI-NIM}. The beam is produced by sending 120\,GeV protons to a two interaction length graphite target. The resulting pions and kaons are focused by a pair of magnetic horns and decay in a 675\,m helium-filled decay pipe to produce neutrinos. The resulting broad-band neutrino energy spectrum peaks at approximately 3~GeV but extends from below 1\,GeV up to 120\,GeV. The horns can be set to focus either positive pions producing a \numu-enhanced beam (forward horn current or FHC) or negative pions producing an \numubar-enhanced beam (reverse horn current or RHC). The analysis reported here includes the dominant FHC \numu and RHC \numubar samples, referred to as the ``focused'' samples, and the minority RHC \numu and FHC \numubar samples, referred to as the ``defocused'' samples. The data in this study were recorded between March 2010 and April 2012 in the ``low-energy'' configuration of the target and horns. The FHC (RHC) data represent $3.175\times10^{20}$ ($1.09\times10^{20}$) protons on target (POT). 

Neutrino interactions in \minerva are reconstructed by first sorting energy deposits into temporally associated groups called ``time slices,'' and then the hits within a time slice are sorted into spatially associated ``clusters" within each scintillator layer. Co-linear clusters within a time slice are used to reconstruct tracks through the tracker and calorimeter regions. Tracks in MINOS are identified in a similar way. Events are required to have one \minerva track matched to a track in MINOS, and that track is identified as coming from a muon. The MINOS-match requirement limits the angular acceptance of muons to within roughly 20$^{\circ}$ of the beam axis and imposes an energy threshold on muon identification of 1.5\,GeV. The momentum of the muon can be determined by either range or curvature in the MINOS detector. To reduce backgrounds, muons are required to have their charge reconstructed with greater than $3.3\sigma$ significance. Each event is assigned a vertex by tracking the muon upstream through the interaction region until no energy is seen in an upstream cone around the track. Events are required to have a reconstructed vertex within the tracker fiducial volume. The recoil energy is reconstructed calorimetrically from clusters in the tracker and downstream calorimeters that are not associated with the muon and are within a 55\,ns window of the muon's clusters. The reconstructed neutrino energy is the sum of the muon and recoil energies\,\cite{MINERvA-NIM}. 

The MINERvA simulation uses the \GENIE package\footnote{\GENIE version 2.6.2 with the GRV98LO parton density functions.} for neutrino interactions\,\cite{GENIE}. The flux is simulated using a \GEANT simulation\footnote{\GEANT version 9.2.p03 with the FTFP physics list.} with corrections to reproduce external hadron production  data\,\cite{NA49a,NA49b,Barton,Denisov,Allaby,Carroll,shine,Lebedev-thesis,tinti-thesis} where the external data are scaled to different incident particle energies using the \FLUKA simulation\,\cite{FLUKA1,FLUKA2}. The version of the flux
used in this analysis was released in 2013 and used in early MINERvA publications\,\cite{minerva-nu-e-flux,Leo-thesis}.
To model the detector reposonse a \GEANT simulation\footnote{Version 9.4.2 with the \textsc{QGSP\_BERT} physics list.} is combined with a model for the readout chain, and the latter is tuned to match the detector response to muons in data.  The statistics of the simulation is a factor of 10 above the data statistics. 

The cross section in bins of \enu is computed as
\begin{equation}
    \sigma(E) = \frac{U(D-B)}{\epsilon\Phi T\times\Delta E},
    \label{eq:cross-section}
\end{equation}
where $D$ is the reconstructed inclusive interaction yield, $B$ is the background yield predicted by the simulation, $U$ is an unfolding operation, $\epsilon$ is the acceptance correction, $\Phi$ is the neutrino flux, $T$ is the number of target nucleons in the fiducial volume, and $\Delta E_\nu$ is the width of the neutrino energy bin. The flux is determined using the \lownu sub-sample of the data:
\begin{equation}
    \Phi = \eta\frac{U_\nu(D_\nu-B_\nu)}{\epsilon_\nu\sigma_\nu T\times\Delta E},
    \label{eq:flux}
\end{equation}
where the subscript $\nu$ indicates that the data, background, acceptance, unfolding, and cross sections have been restricted to $\nu<\nu_0$, which is known as the ``flux'' sample. The small \enu dependence of the \lownu cross section introduces a shape-dependence to the yield of flux sample events apart from the flux, these variations are corrected for by dividing by the \lownu cross section, $\sigma_\nu=\sigma(\nu<\nu_0,E_\nu)$, as predicted, by the \GENIE simulation. The \lownu flux is then normalized such that the extracted inclusive charged-current cross section matches the world average between 9 and 12~GeV. The \lownu normalization factor, described in more detail below, is denoted by $\eta$. 

The event samples used for the flux determinations reported here utilize a $\nu < \nu_0$ cut, with $\nu_0$ varying based on reconstructed neutrino energy. The $\nu_0$ cut choice is a balance between competing concerns. Selecting a larger $\nu_0$ will select more data and yield smaller statistical uncertainties in the flux. Selecting a smaller $\nu_0$ will reduce the energy dependence of the \lownu cross section, and hence the flux-model dependence. As neutrino energy decreases for a given $\nu_0$ cut, the fraction of the inclusive sample that is also in the \lownu flux sample (the overlap fraction) increases, which increases correlations between the \lownu flux and the inclusive cross section. 

Measurements by MINERvA and other experiments show disagreement with \GENIE predictions for both quasielastic scattering\,\cite{K2K-QE,miniboone-quasi,sciboone-qe,minerva-numubarccqe,minerva-numuccqe,minerva-twotrack,minerva-2p2h} and $\Delta$ resonance production\,\cite{MINERvA-pi,miniboone-pi}, which both are largely contained below $\nu=300$\,MeV suggesting a lower limit on the $\nu_0$ cut to reduce sensitivities to mis-modeling. By selecting $\nu_0\geq 300$\,MeV for all neutrino energy regions, discrepancies between data and simulations at low energy transfer should appear as a nearly constant offset in each neutrino energy bin of the flux sample 
(due to their identically restricted final-state phase spaces) 
and are significantly constrained by the normalization technique. Given the competing concerns, the cross section and flux extraction procedure is repeated in parallel for three $\nu_0$ cuts in three energy regions, with the resulting fluxes cross-normalized by independently varying the normalization factors ($\eta$) for the flux in each neutrino energy range. 

\begin{figure}[t]
\includegraphics[height=0.45\linewidth,clip,trim=0.35in 0.15in 1.35in 1.1in]{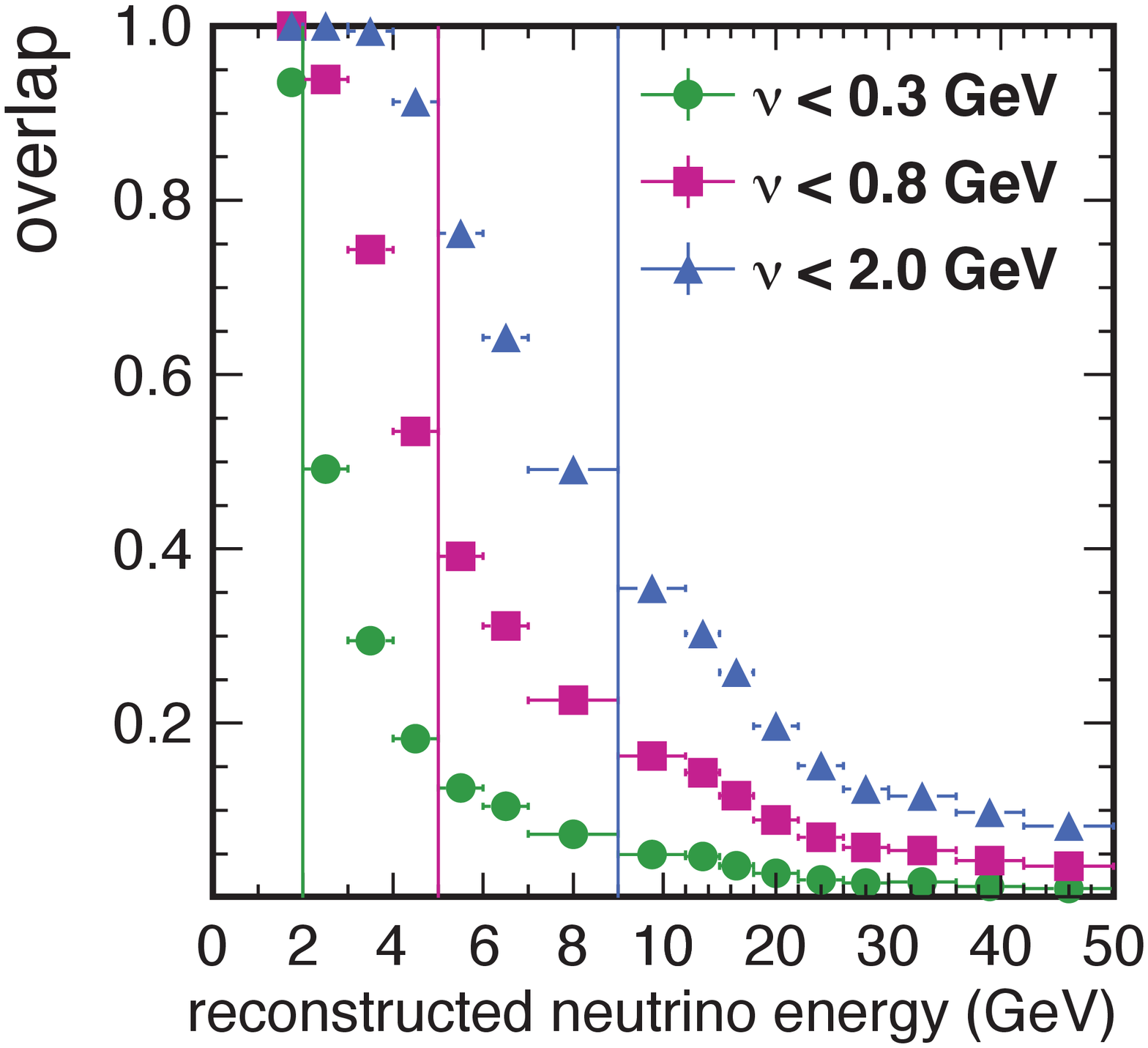}%
\includegraphics[height=0.45\linewidth,clip,trim=0.8in 0.15in  1.35in 1.1in]{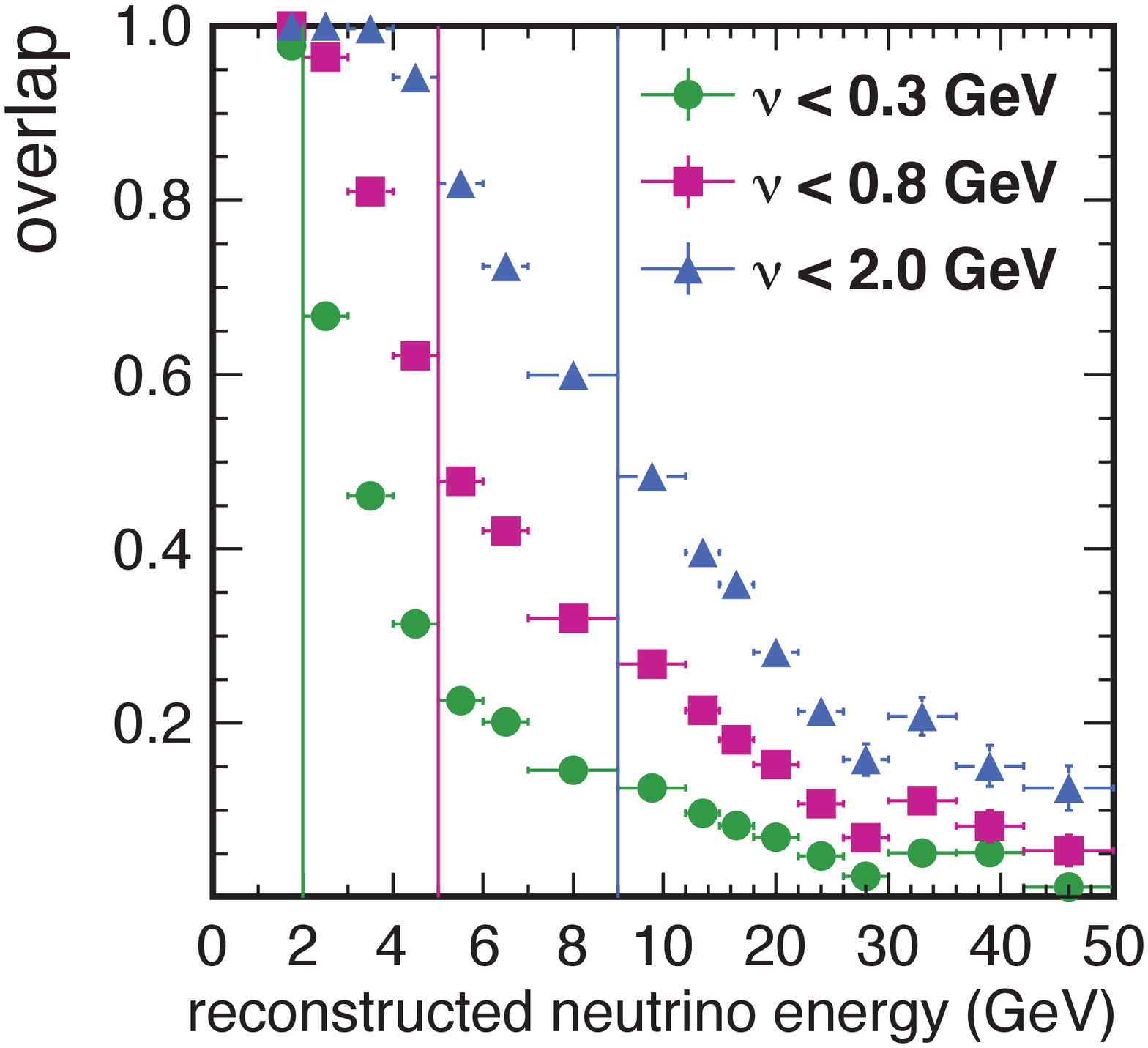}
\caption{The fraction of events in each bin of the \numu (left) and \numubar (right) samples that are included in the \lownu samples (the sample overlap) in the simulation are plotted with their statistical uncertainties. The vertical lines indicate the minimum neutrino energy for each $\nu_0$ cut. Note the bi-linear horizontal energy scale that will be employed throughout this paper.}
\label{fig:overlap}
\end{figure}

\begin{figure}[t]
\includegraphics[height=0.45\linewidth,clip,trim=0 0.1in 1.5in 1.1in]{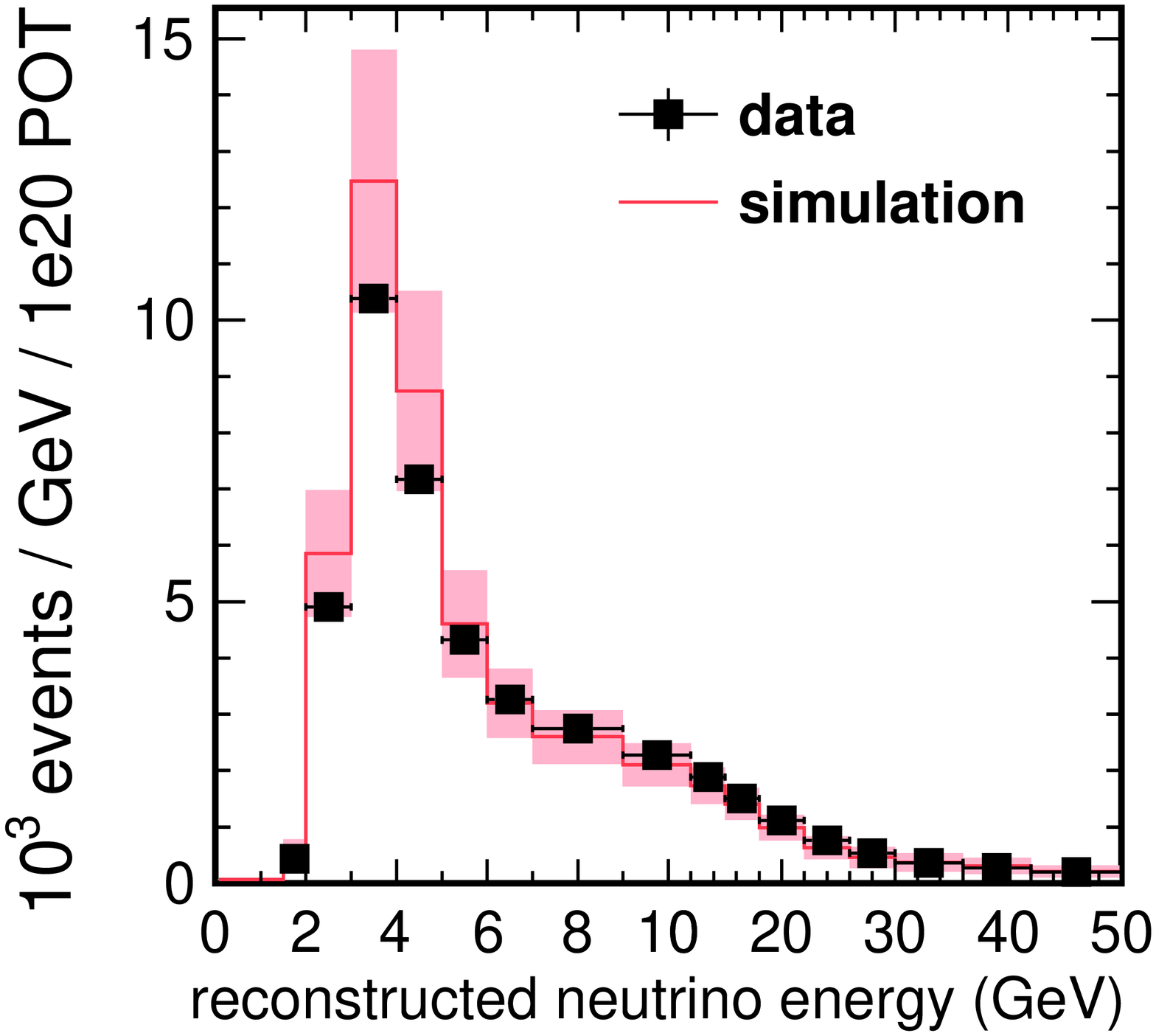}%
\includegraphics[height=0.45\linewidth,clip,trim=0.8in 0.1in 1.35in 1.1in]{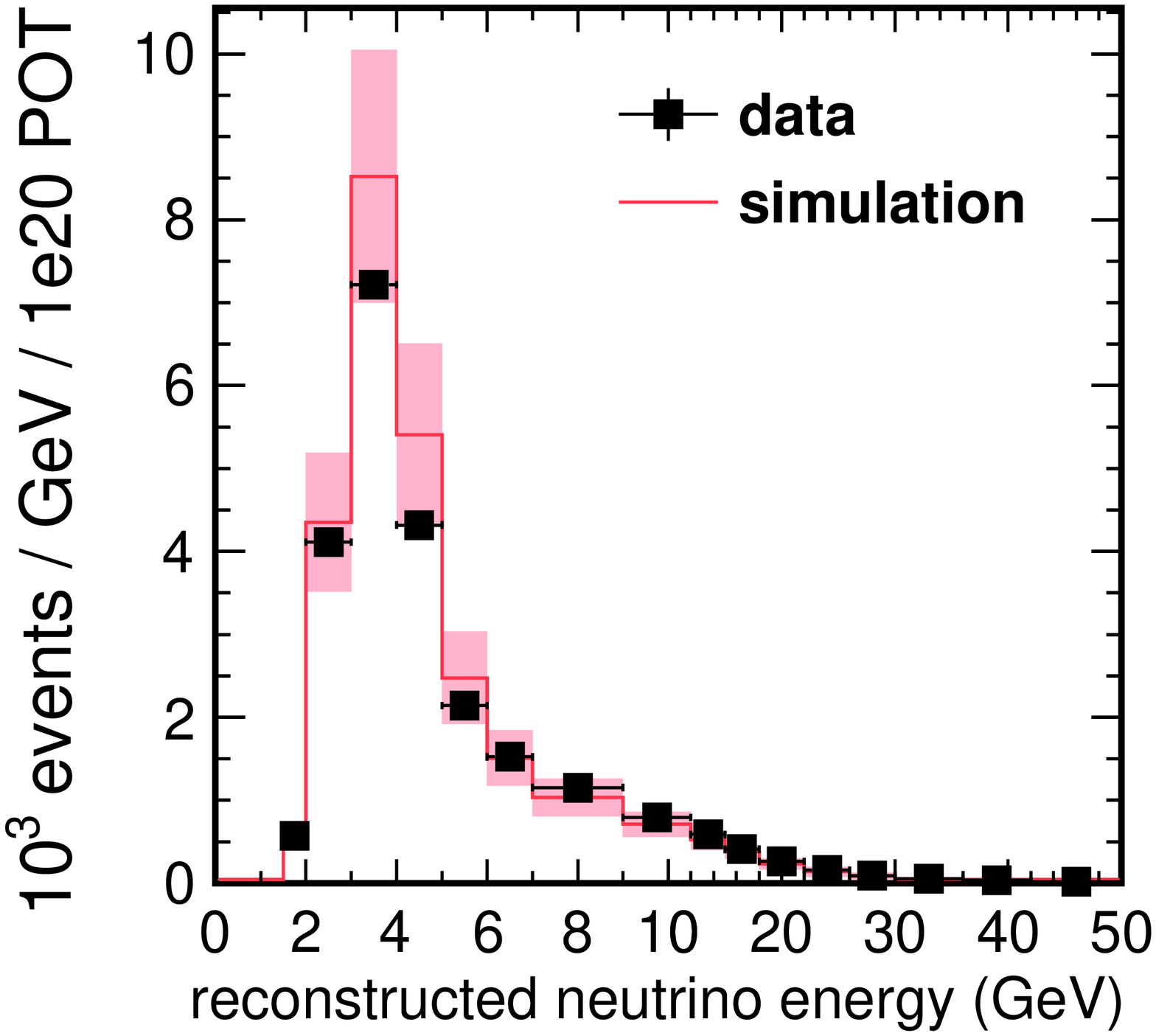}
\includegraphics[height=0.45\linewidth,clip,trim=0 0.1in 1.5in 1.1in]{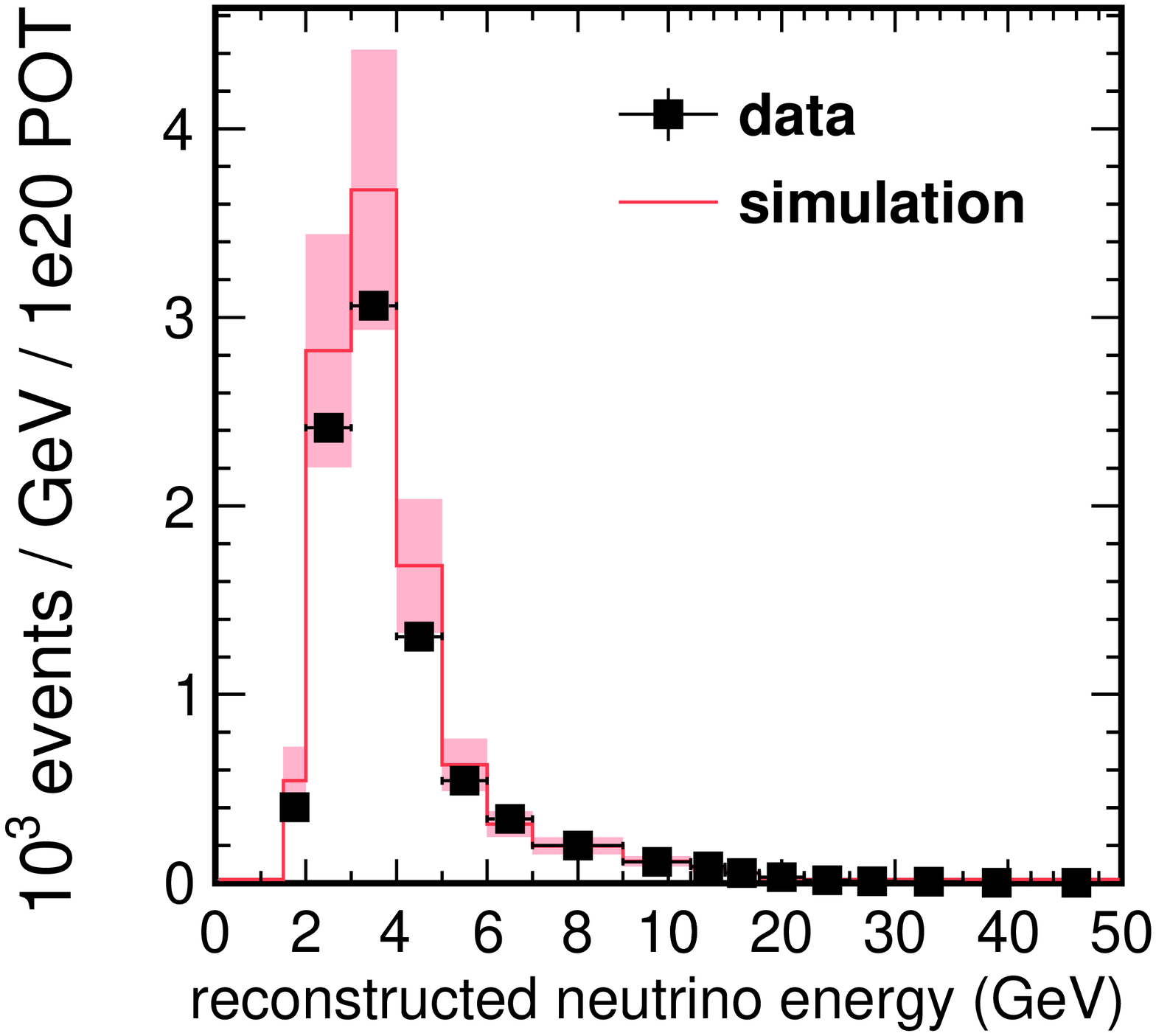}%
\includegraphics[height=0.45\linewidth,clip,trim=0.8in 0.1in 1.35in 1.1in]{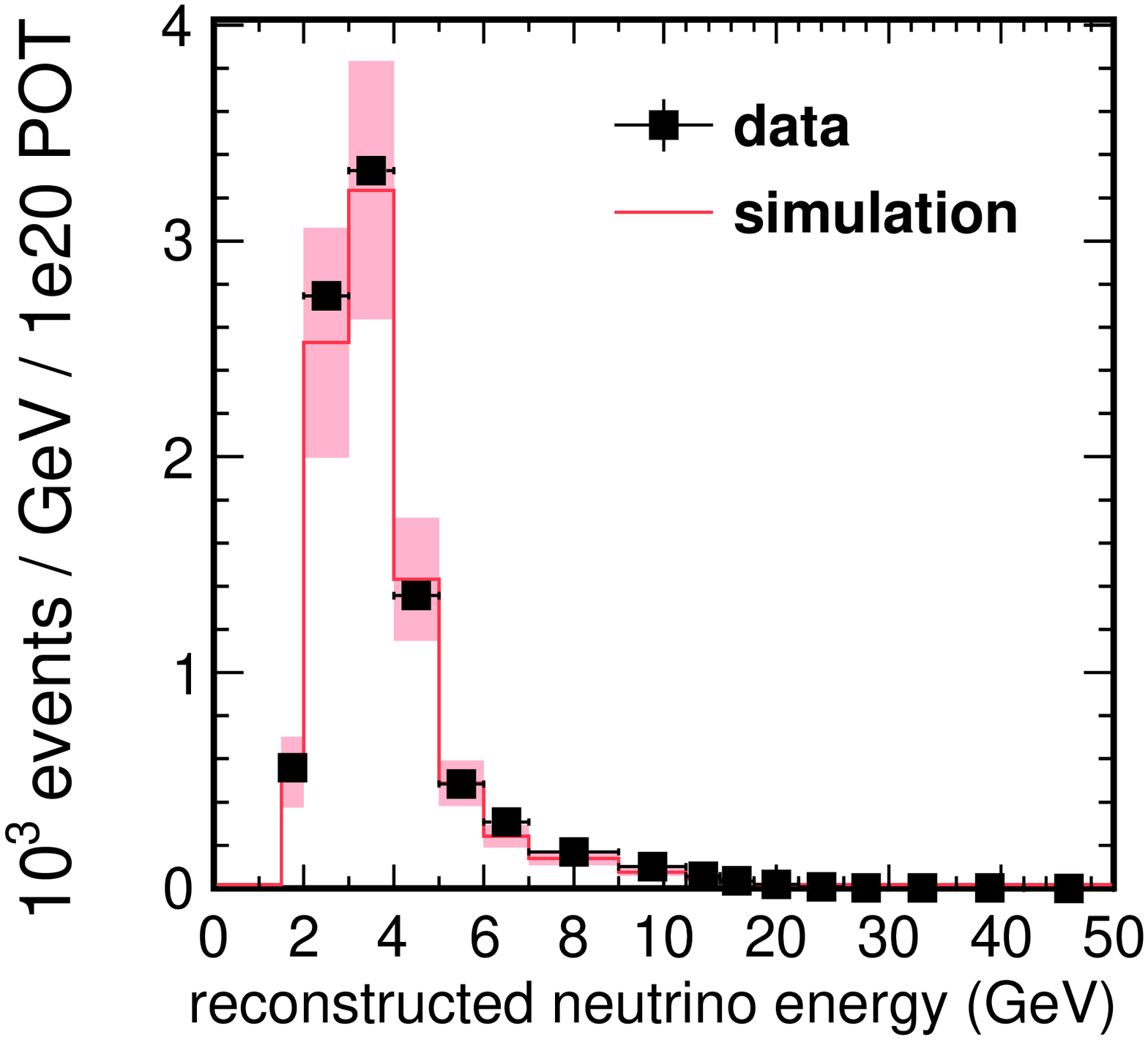}
\caption{Reconstructed focused event yields for neutrinos in FHC (left) and antineutrinos in RHC (right) for the inclusive sample (top) and the sample restricted to $\nu_0=300$\,MeV (bottom) in data and simulation. Data are plotted with statistical uncertainties, while simulated data are plotted with systematic uncertainties.}
\label{fig:events}
\end{figure}

The minimum neutrino energy for each $\nu_0$ cut is set to keep the fraction of cross section events that pass the \lownu cut in each bin less than $50\%$: $\nu_0 = 300$\,MeV for $E_\nu > 2$\,GeV, $\nu_0 = 800$\,MeV for $E_\nu > 5$\,GeV, and $\nu_0 = 2$\,GeV for $E_\nu > 9$\,GeV (c.f. Fig.\,\ref{fig:overlap}). In the lowest \numubar bin this condition is not met with the \lownu sample fraction rising to $67\%$. Fig.\,\ref{fig:events} shows the focused \numu and \numubar yields for the inclusive and $\nu_0=300$\,MeV samples. 

The background in the selected samples is dominated by muon charge misreconstruction, and the background fraction is determined from the simulation. This fraction is negligible for the focused samples, but up to 5\% for the defocused samples. The simulated yield (reconstructed signal plus background) is scaled to the data in each energy bin before the background subtraction. 

The data after background subtraction are unfolded to true neutrino energy to remove detector resolution and biases in muon and recoil energy reconstruction using two-iteration Bayesian unfolding\,\cite{Bayes-unfold}. The migration probabilities are summarized in matrices of true versus reconstructed neutrino energy that are evaluated independently for each sample using the simulation. The neutrino energy binning is chosen to keep the bin migration well below 50\% to control the bin-to-bin correlations in the unfolded distributions. 

\begin{figure}[t]
\includegraphics[height=0.45\linewidth,clip,trim=0 0.15in 1.0in 1.1in]{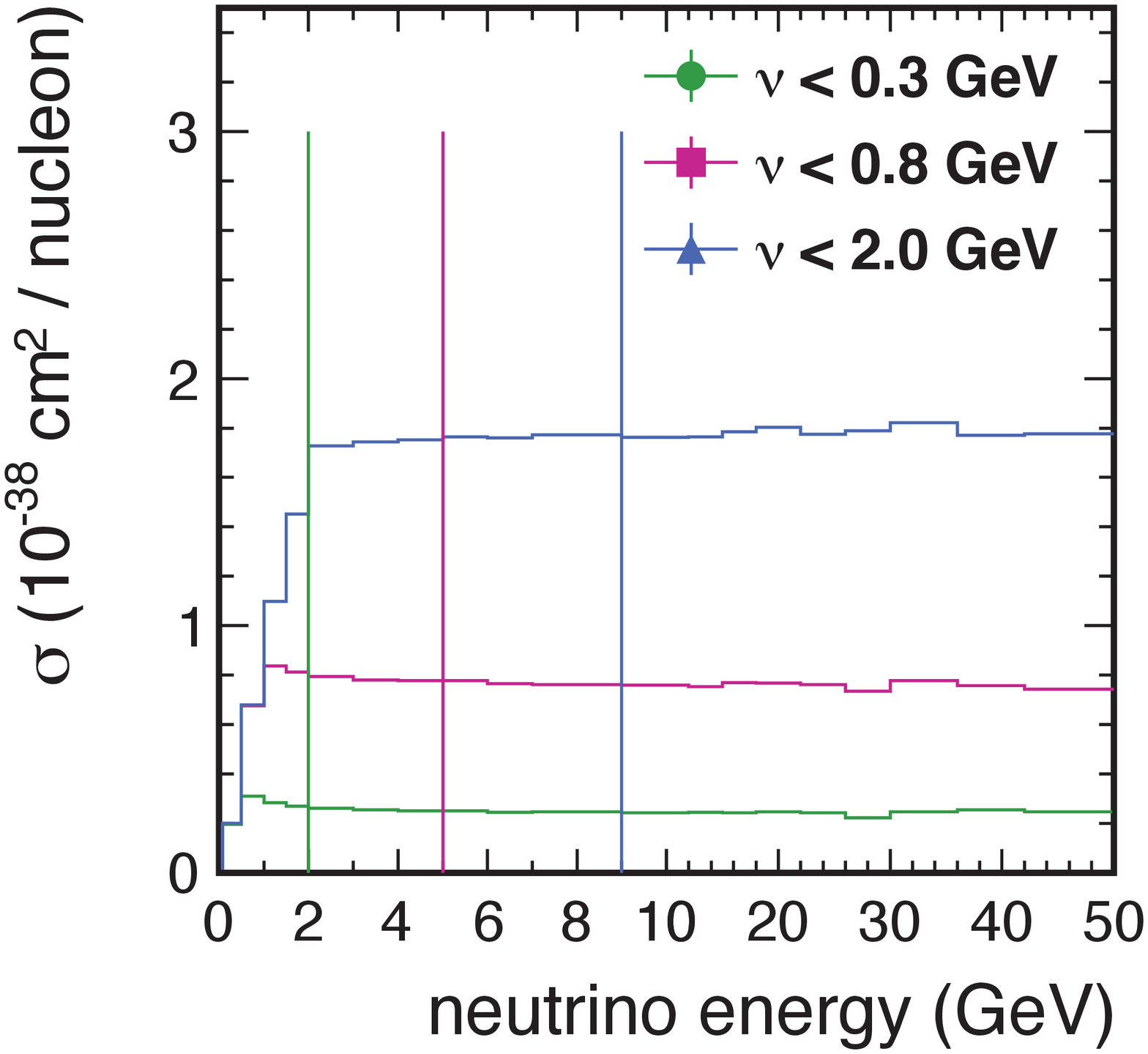}%
\includegraphics[height=0.45\linewidth,clip,trim=0.8in 0.15in 1.0in 1.1in]{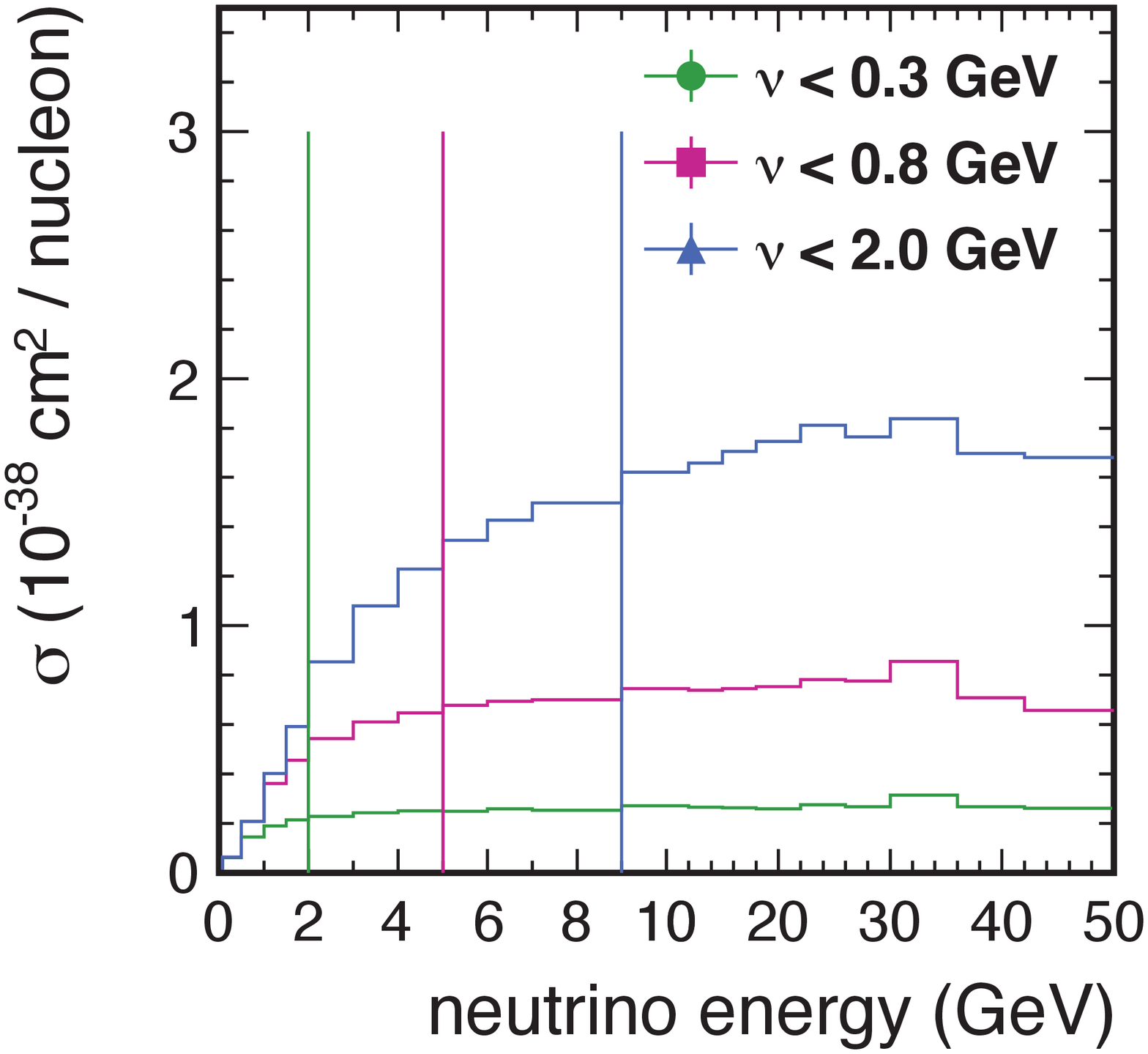}
\caption{\LowNu cross sections for neutrinos and antineutrinos as derived from the \GENIE event generator (version 2.6.2). Vertical lines indicate the minimum neutrino energy for each $\nu_0$ cut.}
\label{fig:GENIE-xsec}
\end{figure}

The acceptance correction is derived from simulation and primarily accounts for reconstruction inefficiency due to the MINOS track-matching requirement. The acceptance also corrects for migration into, or out of, the fiducial volume and migration across the $\nu_0$ cut in the flux samples. In the \lownu samples, where most of the neutrino energy is transferred to the muon, the acceptance is greater (rising to 70\% at 5\,GeV and 90\% at 20\,GeV) than in the inclusive samples (rising to 40\% at 5\,GeV and 60\% at 20\,GeV). The energy dependence of \signu arising from both the non-zero value of $\nu_0$ and the small $Q^2$ dependence of the structure functions is evaluated utilizing the \GENIE neutrino interaction simulation (Fig.\,\ref{fig:GENIE-xsec}). 

The normalization of the extracted flux (Eq.\,\ref{eq:flux}) is performed successively for each $\nu_0$ cut yielding a set of three normalizations. The $\nu_0=2$\,GeV flux sample is normalized such that the extracted total cross section matches an externally measured value in the 9--12 GeV bin. The normalization energy is selected to be high enough that $\sigma(E_\nu)/E_\nu$ is approximately constant but low enough to mitigate statistical uncertainties. The $\nu_0 = 800$\,MeV result is then cross-normalized to the $\nu_0 = 2$\,GeV result to minimize the $\chi^2$ difference for 9\,GeV $<$ $E_\nu$ $<$ 22\,GeV. The $\nu_0 = 300$\,MeV result is then cross-normalized to the $\nu_0$ = 800\,MeV results for 5\,GeV $<$ $E_\nu$ $<$ 22\,GeV. 

To compare these cross sections to prior measurements, which are on different nuclei, we report isoscalar-corrected cross sections by multiplying by the ratio of the per nucleon cross sections on C$^{12}$ and the mix of materials in \minerva as derived from the \GENIE simulation. The isoscalar correction for neutrinos varies from 0.96 at low energies to 0.98 at high energies, while the correction for antineutrinos is essentially constant at 1.03. 

The systematic uncertainties on the shapes of the flux and cross section measurements originate from a variety of sources. The MINOS collaboration reports a 2\% uncertainty on range-based momentum measurements\,\cite{MINOS-NIM}. The systematic uncertainty of a curvature-based measurement is driven by the uncertainty on the magnetic field model of the detector with uncertainties of 0.6\% for $p_\mu>1$\,GeV/c and 2.5\% for $p_\mu<1$\,GeV/c, which is added in quadrature to the range-based uncertainty\,\cite{MINERvA-NIM}. The simulated efficiency for the muon reconstruction is benchmarked by extrapolating reconstructed tracks in each detector and measuring the rate of failures in the other detector. 

The systematic uncertainty on the calorimetric reconstruction of $\nu$ is estimated by convolving individual particle uncertainties into the \GENIE model's ensemble of final states. The uncertainties on the proton (10\%) and meson (5\%) calorimetric responses are derived from test beam studies.\footnote{Note this analysis uses an older version of the simulation than Ref.\,\cite{TB-NIM} and does not incorporate the improvements derived from the test beam program which would have reduced these uncertainties.} The uncertainty on the electromagnetic ($e^{\pm}, \pi^0, \gamma$) response (3\%) is derived from studies of Michel electrons\,\cite{MINERvA-NIM}, the $\pi^0$ mass peak\,\cite{MINERvA-pi0}, and a test beam study\,\cite{TB-NIM}. The neutron response uncertainty (15\%) is based on comparisons between measured inelastic cross sections of neutrons on carbon and \GEANT simulations of those measurements. At low $\nu$, where QE final states dominate, the uncertainty approaches 10\% (15\%) for neutrinos (antineutrinos). At high $\nu$, where the majority of the recoil energy is carried by charged and neutral pions, the uncertainty approaches 6\% for both neutrinos and antineutrinos. 

Biases in the recoil energy measurement can come from both accidental activity in the detector and from muon bremsstrahlung.  Because the recoil energy measurement comes from summing all the energy in the detector within a 55\,ns time window, energy from other neutrino interactions coincident with the neutrino interaction of interest will be added to the recoil energy.  Furthermore, the primary muon in the event may undergo large energy loss when the muon produces knock-on electrons or emits a photon via bremsstrahlung. These secondary particles travel some distance from the muon's trajectory and may not be associated with the muon during reconstruction, which again leads to an overestimate of the recoil energy. 

Accidental activity is modeled by overlaying a single simulated neutrino interaction onto a randomly selected data spill. Then, we compare the off-track energy for muon tracks between data and a simulation that includes accidental activity. The small data-simulation difference is added as a correction to the recoil energy simulation.  

The \lownu flux shape measurement is largely insensitive to the neutrino flux used in the simulation. Uncertainties in the shape of this simulated flux, however, could subtly affect the unfolding correction. The flux uncertainty is dominated by the uncertainty in the external hadron-production measurements\,\cite{Leo-thesis}. There are also uncertainties in modeling the beamline focusing system\,\cite{zarko-thesis}. For the portion of the flux produced by interactions on nuclei that are not covered by external data (e.g. the aluminum horns and steel decay-pipe walls) the systematic uncertainties are derived from the spread between different hadronic interaction models\,\cite{Leo-thesis}. The effect of these uncertainties in the initial flux are found to be negligible, as verified by a data-driven test described later in the paper. 


The uncertainties in the \GENIE simulation are evaluated based on recommendations from the GENIE Collaboration and incorporate uncertainties in the external data used to tune their models\,\cite{GENIE-PUM}. The most significant uncertainties in the \lownu flux shape measurement, and hence the cross section measurement, are related to QE scattering and resonance production, which dominate the \lownu samples. 

\begin{figure}[t]
\includegraphics[height=0.47\linewidth,clip,trim=0in 0.1in 1.45in 1.in]{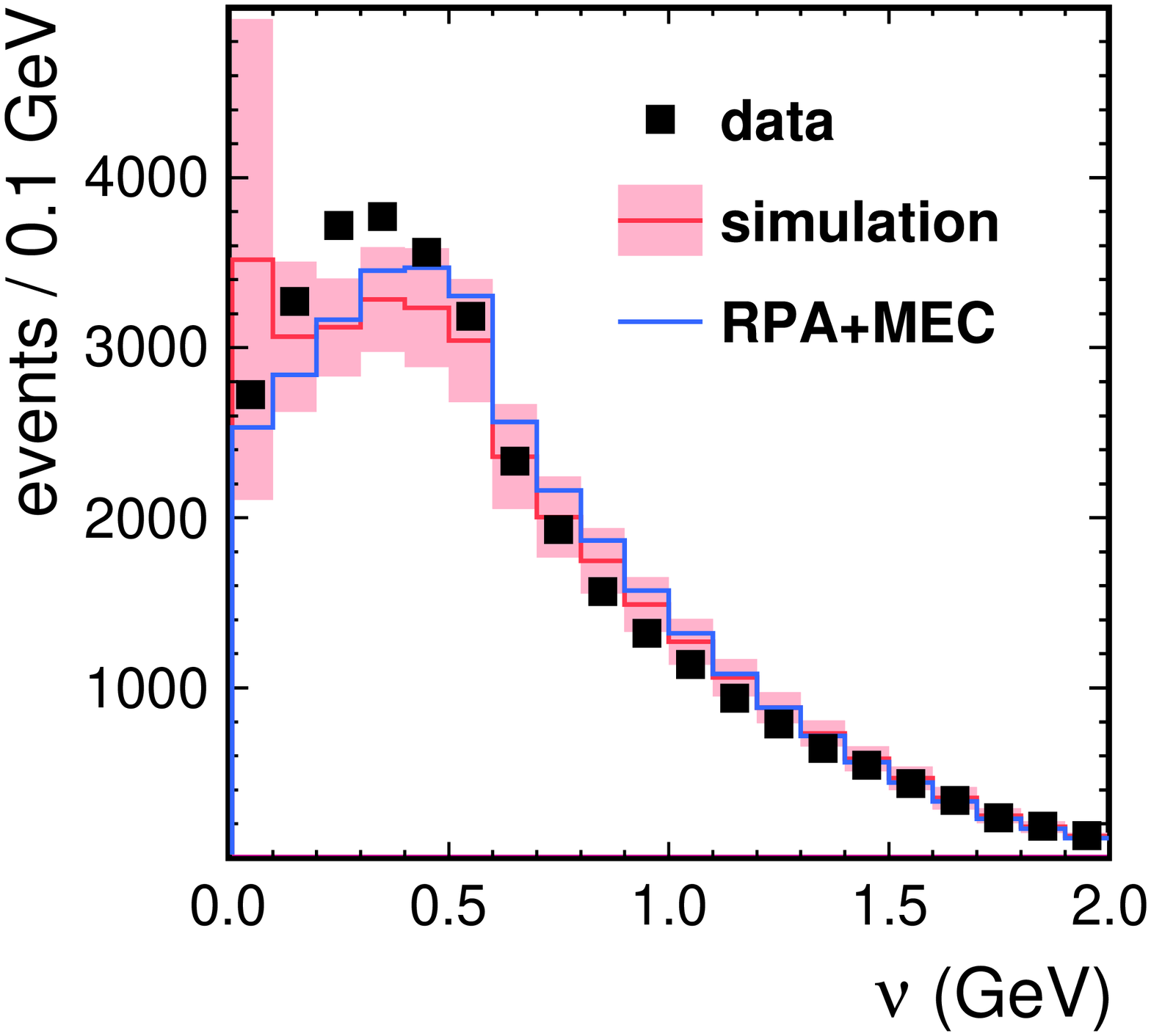}%
\includegraphics[height=0.47\linewidth,clip,trim=0.7in 0.1in 1in 1.in]{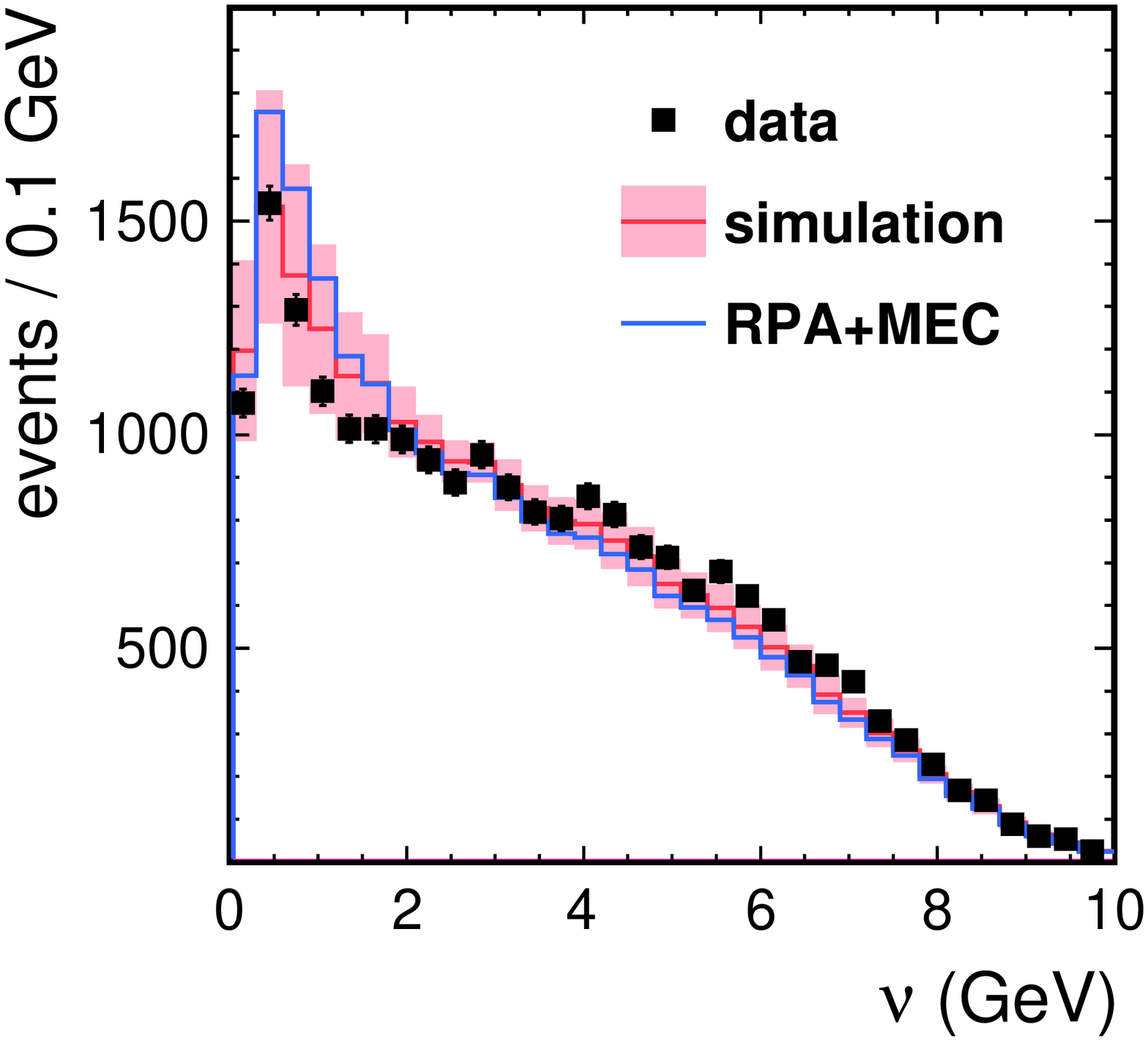}
\caption{Area-normalized event yields as a function of $\nu$ for \numu events with $3<E_\nu<4$\,GeV (left) and $9<E_\nu<12$\,GeV (right). The square points are data plotted with statistical uncertainties, the red line is the default \GENIE prediction with shape-only systematic uncertainties, and the blue line is the modified prediction including RPA and MEC effects.} 
\label{fig:nu-dist}
\end{figure}

There are two important nuclear effects that have not yet been incorporated into the GENIE version used for this analysis.  The random phase approximation (RPA) model incorporates long-range nucleon-nucleon correlations resulting in a reduced inclusive cross section at low energy transfers. The meson exchange current (MEC) model describes a process in which nucleons within a nucleus exchange momentum via a pion during the interaction resulting in two-nucleon final states and an increased cross section in the kinematics between the QE and $\Delta$ production processes. Since the default \GENIE model does not simulate these processes, a \GENIE implementation\,\cite{genie-velencia} based on the Valencia model~\cite{valencia-model} was employed to evaluate their effects on the QE process, which include a suppression for $\nu$ $<$ 200\,MeV and modest enhancement at higher $\nu$ (Fig.\,\ref{fig:nu-dist}). The agreement between the neutrino data and simulation is significantly improved with the RPA and MEC models\,\cite{minerva-2p2h}. The entire difference between the \GENIE default prediction and the prediction including the RPA and MEC models is taken as the systematic uncertainty for these nuclear effects. The same uncertainty is also applied to the antineutrino analysis. 

The \GENIE Collaboration recommends a large uncertainty on the charged-current quasi-elastic axial mass ($M_A^{CCQE}$) to cover observed discrepancies in experimental data\,\cite{GENIE-PUM}. This large range is intended to cover unmodeled effects like the those in the RPA and MEC models. Since those uncertainties are directly assessed, as described above, the large $M_A^{CCQE}$ variations are not included in the systematic uncertainty of the interaction model. Instead a smaller estimated (2\%) variation, resulting from an analysis of exclusive-state measurements from precision measurements\,\cite{Bodek-ma}, is evaluated and has a negligible effect. 

This \lownu-flux analysis results in the shape of the neutrino fluxes that are then used to determine the shapes of the charged-current inclusive cross sections for neutrinos and antineutrinos. The fluxes are normalized to reproduce inclusive total cross sections between $9<E_\nu<12$\,GeV. The \numu inclusive cross section normalization is taken from the NOMAD experiment\,\cite{NOMAD-xsec}, which reports a cross section measurement with 3.6\% systematic uncertainty between $9<E_\nu<12$\,GeV. No correspondingly precise single measurement exists for \numubar. The \numubar normalization is instead set to the predictions from the \GENIE simulation, which is tuned to the world's inclusive data. The uncertainty is derived according to the prescribed \GENIE model parameter uncertainties and evaluates to 10.6\%. The normalization process adds a flat uncertainty that is the quadrature sum of the external uncertainty and the statistical uncertainty in the normalization bin. The resulting normalization factors ($\eta$ from Eq.\,\ref{eq:flux}) are listed in Table\,\ref{tbl:norm} for each sample. A value of $\eta < 1$ indicates that data prefer a lower \lownu cross section than modeled in the \GENIE simulation. 

\ifnum\DoPrePrint=1
\begin{table}[t]
   \setlength{\tabcolsep}{4pt}
\begin{tabular}{|c|cc|cc|}
\hline
$\nu_0$ cut   &  \multicolumn{2}{|c|}{neutrinos}   & \multicolumn{2}{|c|}{antineutrinos}         \\
(GeV)     &  $\eta$ & Yield ($10^3$ events)& $\eta$ &  Yields ($10^3$ events)  \\
\hline
2.0 & $0.925 \pm 0.009$ & 119.1 & $0.943 \pm 0.021$ & 24.9 \\
0.8 & $0.958 \pm 0.009$ & 75.4 &$1.085 \pm 0.019$ & 18.2 \\
0.3 & $0.946 \pm 0.012$ & 29.6 &$1.200 \pm 0.020$ & 10.4 \\
Incl. &  n/a & 215.2 & n/a & 32.7 \\
\hline
\end{tabular}
\caption{Normalization factors ($\eta$) and their statistical uncertainties for neutrinos and antineutrinos by $\nu_0$ sample and the total event yields. The uniform 3.6\% (10.6\%) external normalization uncertainty for \numu (\numubar), described in the text, are in addition to the uncertainties in this table.}
\label{tbl:norm}
\end{table}
\else
\begin{table}[t]
   \setlength{\tabcolsep}{4pt}
\begin{tabular}{lccr}
\hline
&\multirow{ 2}{*}{$\nu_0$ cut (GeV)}  &  \multirow{ 2}{*}{$\eta$} & \multicolumn{1}{c} {Yield} \\
&   &   &\multicolumn{1}{c}{($10^3$ events)} \\
\hline
\multirow{ 4}{*}{Neutrinos}& 2.0 & $0.925 \pm 0.009$ & 119.1~~~~~ \\
&0.8 & $0.958 \pm 0.009$ & 75.4~~~~~ \\
&0.3 & $0.946 \pm 0.012$ & 29.6~~~~~ \\
&Inc. &  n/a & 215.2~~~~~ \\
\hline
\multirow{ 4}{*}{Antineutrinos}&2.0 &  $0.943 \pm 0.021$ & 24.9~~~~~ \\
&0.8 & $1.085 \pm 0.019$ & 18.2~~~~~ \\
&0.3 & $1.200 \pm 0.020$ & 10.4~~~~~ \\
&Inc. &  n/a & 32.7~~~~~ \\
\hline
\end{tabular}
\caption{Normalization factors ($\eta$) and their statistical uncertainties for neutrinos and antineutrinos by $\nu_0$ sample and the total event yields. The uniform 3.6\% (10.6\%) external normalization uncertainty for \numu (\numubar), described in the text, are in addition to the uncertainties in this table.}
\label{tbl:norm}
\end{table}

\fi

\begin{figure}[t]
\includegraphics[height=1.4in,clip,trim=0.1in 1.15in 1.0in 1in]{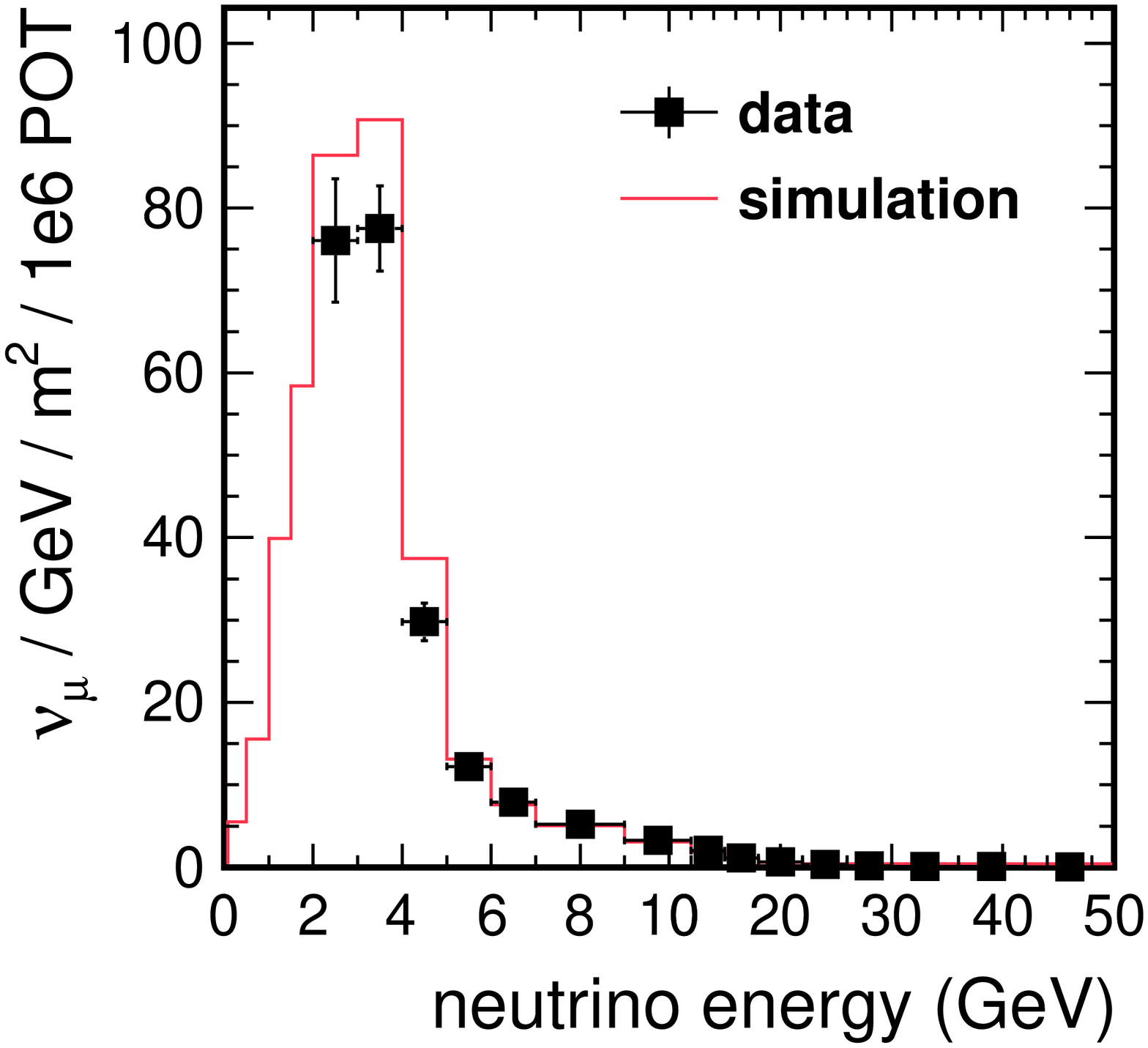}%
\includegraphics[height=1.4in,clip,trim=1.2in 1.15in 1.0in 1in]{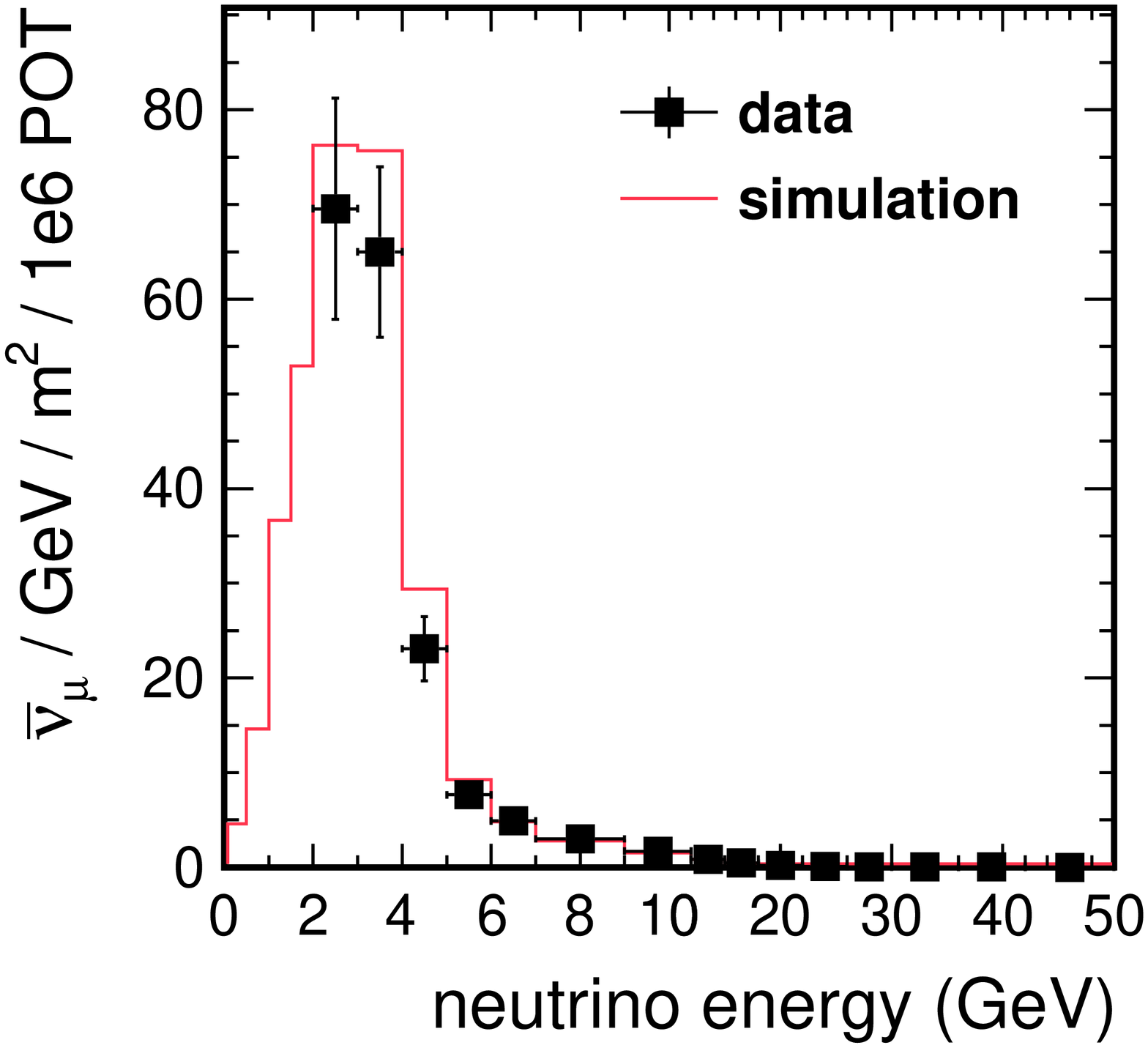}
\includegraphics[height=1.63in,clip,trim=0.1in 0.10in  1.0in 1in]{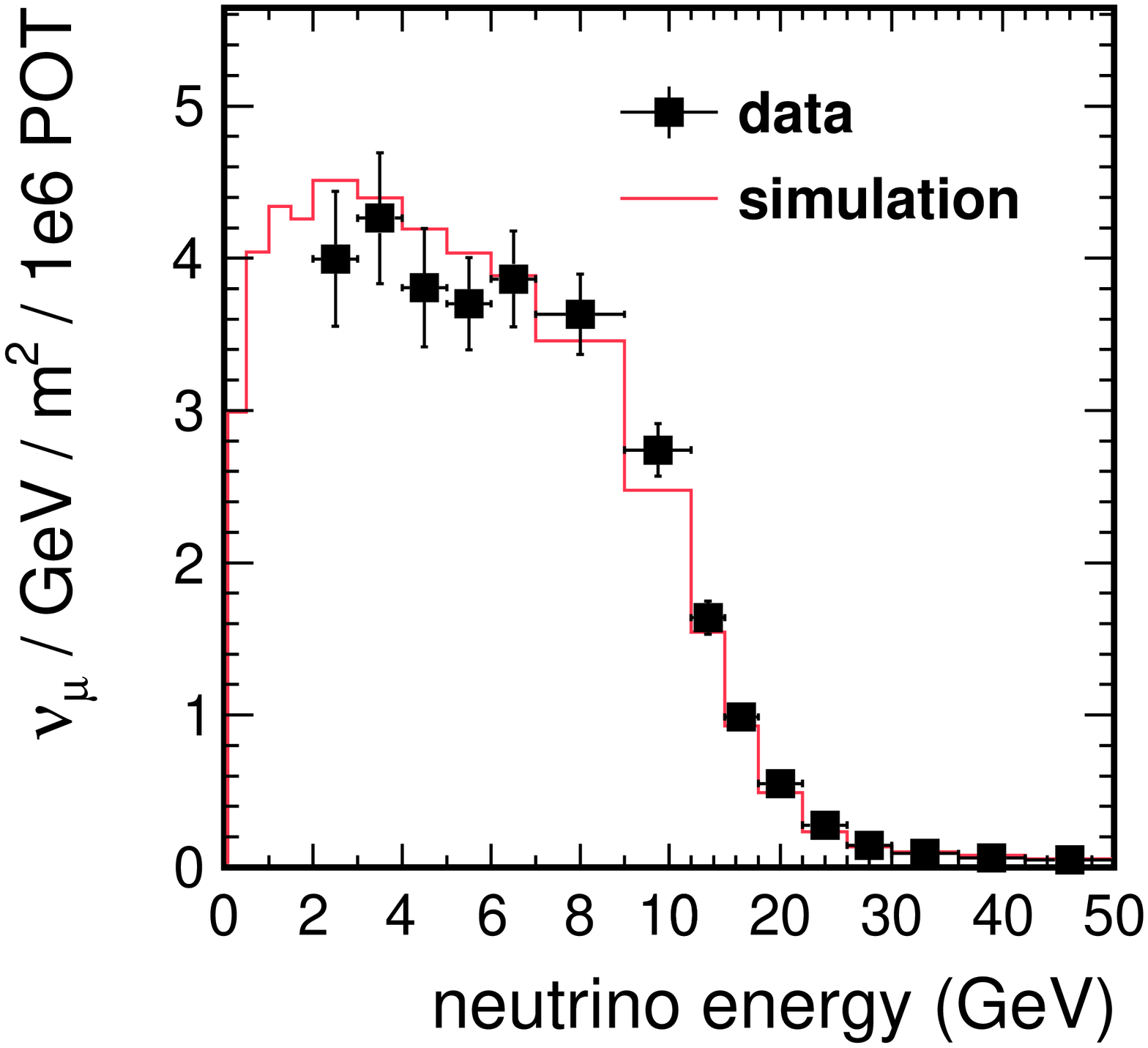}%
\includegraphics[height=1.63in,clip,trim=1.2in 0.10in  1.0in 1in]{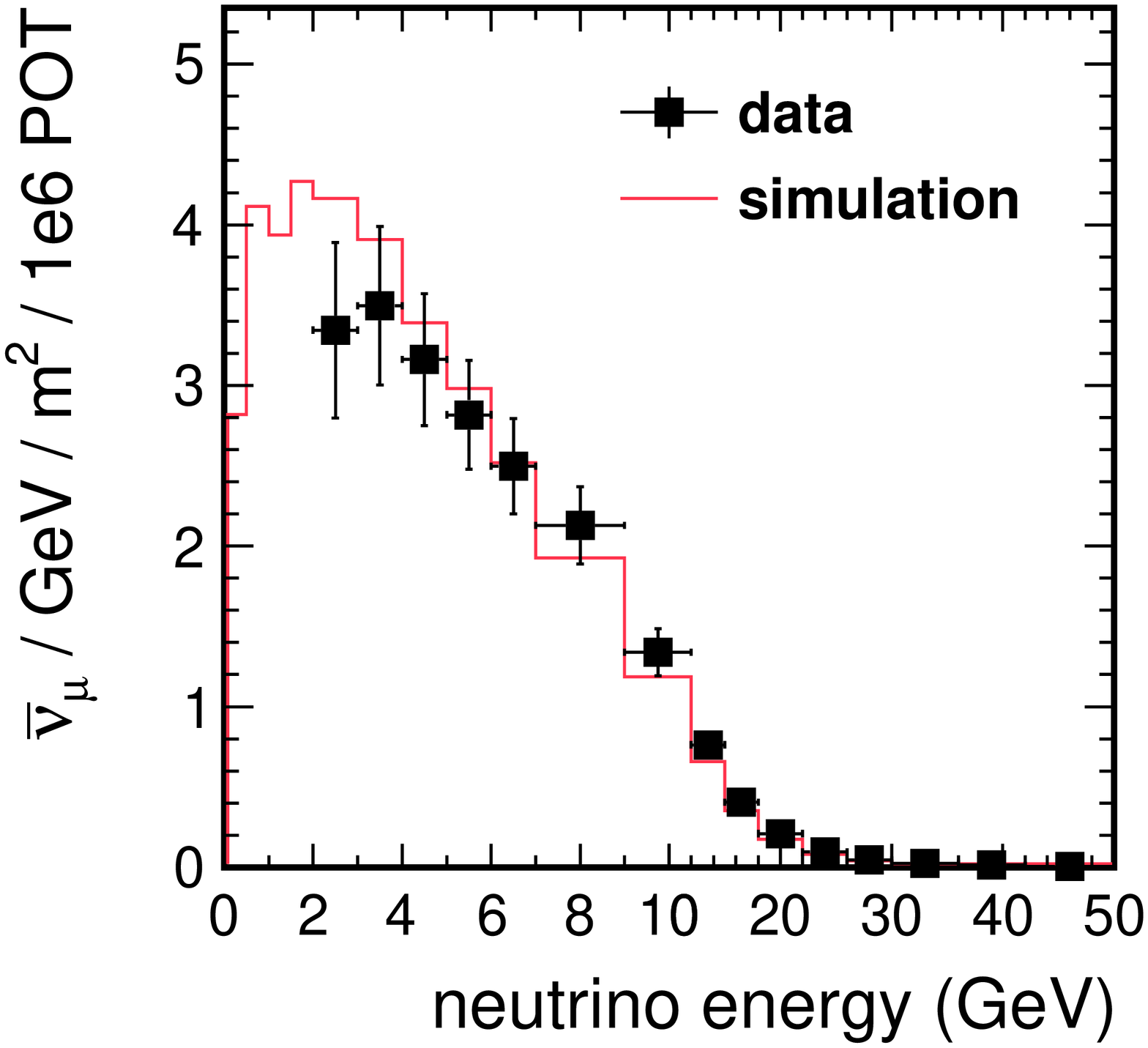}
\caption{Measured fluxes for neutrinos in the FHC tune (top, left), antineutrinos in the RHC tune (top, right), neutrinos in the RHC tune (lower left), and antineutrinos in the FHC tune (lower right) are shown with combined statistical and systematic uncertainties and are compared to the {\em a priori} flux calculation used in the simulation.}
\label{fig:flux}
\end{figure}

The resulting normalized neutrino and antineutrino fluxes are shown in Fig.\,\ref{fig:flux} for both the FHC and the RHC beam configurations. The external normalization uncertainty dominates over the shape-dependent uncertainties\footnote{Tables with error breakdowns and covariance matrices for the extracted fluxes are included in supplementary materials.}. Muon energy reconstruction also contributes a significant uncertainty at lower neutrino energies as the measured shape of the flux is determined from \lownu events where the majority of the energy goes into the final-state muon. Each of the flux results show a similar deficit in lower energies (\mbox{2--5\,GeV}) and a modest excess at higher energies when compared to {\em a priori} flux estimations. 

An independent {\em in situ} flux constraint can be obtained by measuring neutrino elastic scattering off atomic electrons. A purely leptonic process, the cross section is precisely calculable, allowing a measurement of the integrated neutrino flux based on the observed interaction rate. The MINERvA electron-scattering analysis\,\cite{minerva-nu-e-flux} agrees better (at the 10\% level) with the fluxes measured by this analysis than the version of the flux used in the simulation\,\cite{Leo-thesis}.

\begin{figure}[t]
\centering
\ifnum\DoPrePrint=1
\includegraphics[width=.5\textwidth,clip,trim=0.25in 0.7in 1.2in 1.1in]{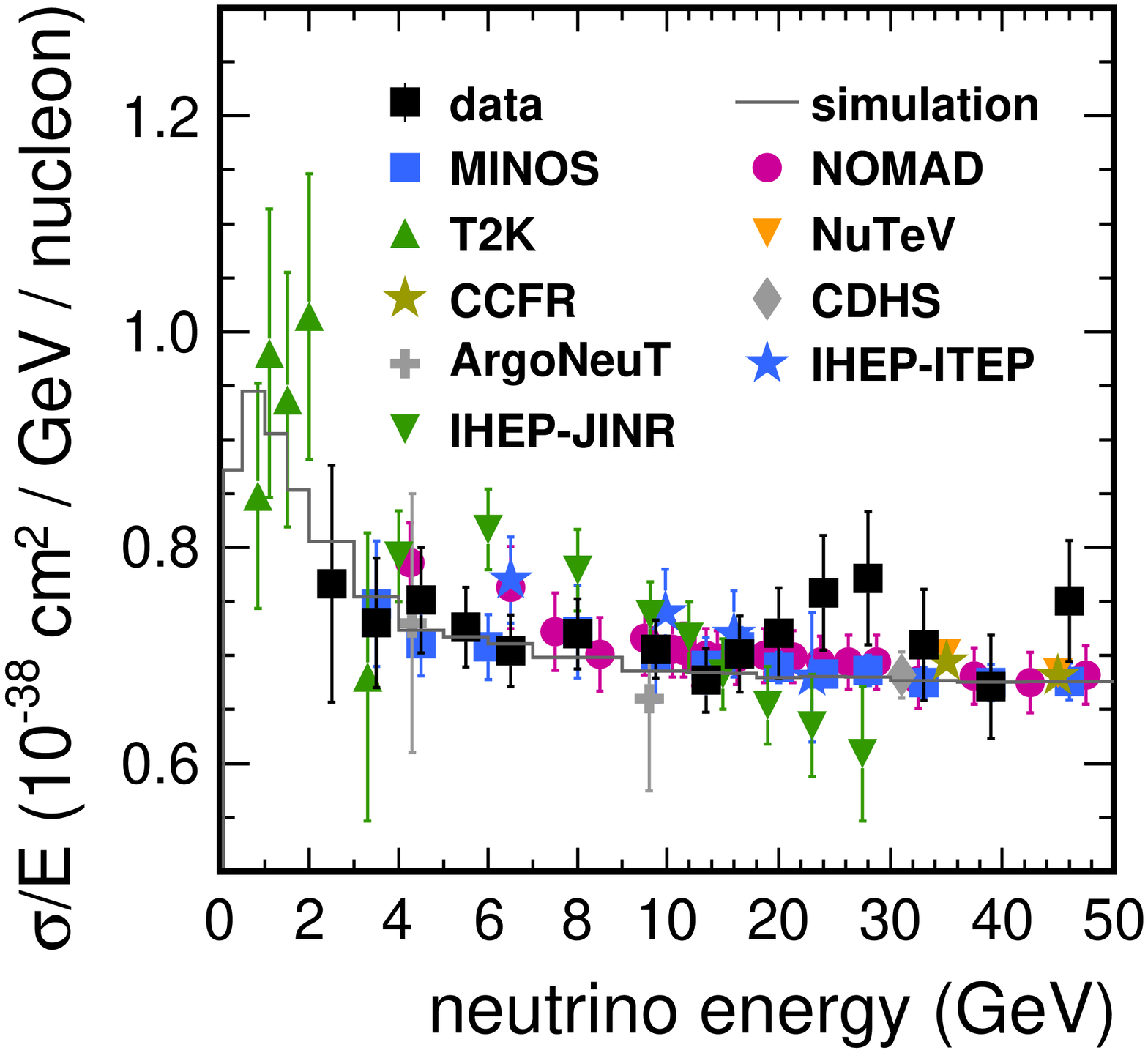}%
\includegraphics[width=.46\textwidth,clip,trim=1.0in 0.7in 1.2in 1.1in]{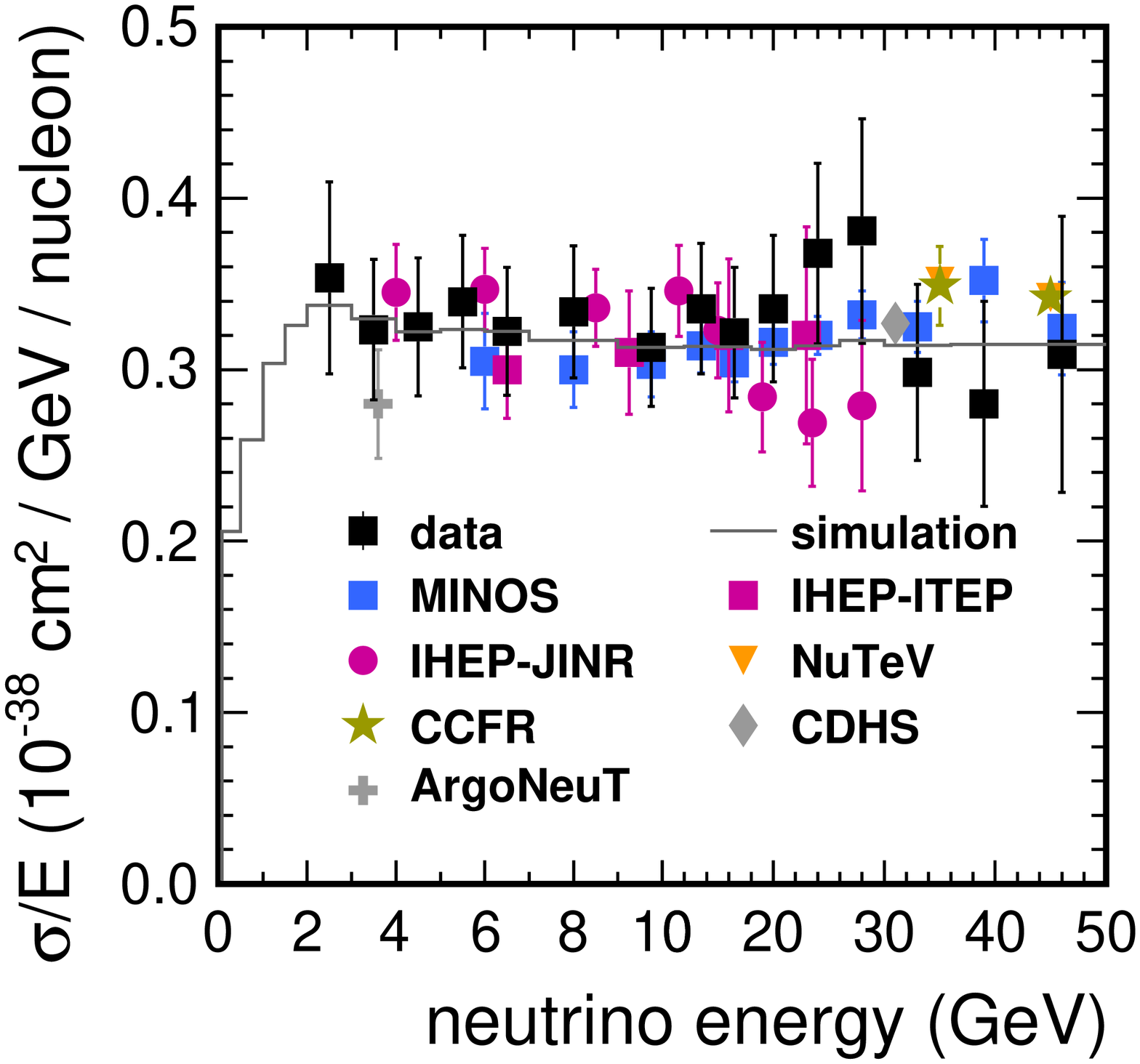}
\else
\includegraphics[width=.25\textwidth,clip,trim=0.25in 0.0in 1.2in 1.1in]{66_nomad_xsec_nu_color.eps}%
\includegraphics[width=.227\textwidth,clip,trim=1.0in 0.0in 1.2in 1.1in]{60_iso_xsec_anti_color.eps}
\fi
\caption{Measured isoscalar-corrected inclusive charged-current neutrino cross section for neutrinos (left) and antineutrinos (right) plotted with combined statistical and systematic uncertainties and compared to world data and the GENIE simulation.} 
\label{fig:xsec}
\end{figure}

Figure \ref{fig:xsec} shows the isoscalar-corrected inclusive charged-current cross sections for both \numu and \numubar with combined statistical and systematic uncertainties\footnote{Tables with error breakdowns and covariance matrices for the extracted cross sections are included in supplementary materials.}. By definition, the systematic uncertainties collapse to the normalization uncertainty in the 9--12\,GeV bin. The \numu cross section uncertainty is dominated by the normalization, except in the lowest energy bin where the \GENIE model and the calorimetric reconstruction each contribute significantly to the uncertainties. The \numubar cross section uncertainty is dominated by the normalization uncertainty, except at high energies where the statistical uncertainty is significant. 

As a cross-check the cross section analysis is also performed using the ``defocused'' samples (neutrinos in RHC and antineutrinos in FHC). The polarity of the MINOS magnet is set to defocus the muons in these samples resulting in a significantly different acceptance correction. In the focused samples the backgrounds are negligible while in the defocused samples there are substantial backgrounds from charge-misidentified muons in MINOS. The extracted cross sections from the defocused and focused samples agree to within their statistical uncertainties. Another cross check is to repeat the analysis starting with the extracted \lownu fluxes instead of the simulated fluxes to check for consistency. The resulting cross sections are consistent to better than 1\% in every bin, which verifies that the extracted cross sections are insensitive to the initially assumed simulated fluxes. 

Figure \ref{fig:xsec} compares the measured isoscalar-corrected charged-current inclusive \numu cross section to the MINOS\,\cite{MINOS-xsec}, T2K\,\cite{T2K-xsec1,T2K-xsec2, T2K-xsec3}, CCFR\,\cite{CCFR-xsec}, Argoneut\,\cite{Argoneut-xsec1,Argoneut-xsec2} IHEP-JINR\,\cite{IHEP-JINR-xsec}, NOMAD\,\cite{NOMAD-xsec},  NuTeV\,\cite{NuTeV-xsec}, CDHS\,\cite{CDHS-xsec}, and IHEP-ITEP\,\cite{IHEP-ITEP-xsec} measurements. MINOS measured inclusive cross sections on iron using the \lownu flux technique. T2K measured the cross sections in the ND280 and INGRID detectors. The T2K flux was determined from a simulation of their beamline constrained by particle production data. For this comparison, the T2K results were isoscalar corrected with a value of 1.043 for the scintillator data point (0.85\,GeV), and a value of 0.98 for the iron data points (1.1, 1.5, 2.0, and 3.3\,GeV) derived from the MINOS correction\,\cite{Debdatta-thesis}. The results of this analysis agree with the NOMAD and MINOS results within uncertainties, but extend to lower energies. 

Figure\,\,\ref{fig:xsec} also compares the measured isoscalar-corrected charged-current \numubar inclusive cross section to the MINOS\,\cite{MINOS-xsec}, IHEP-ITEP\,\cite{IHEP-ITEP-xsec}, and IHEP-JINR\,\cite{IHEP-JINR-xsec}, CCFR\cite{CCFR-xsec}, Argoneut\,\cite{Argoneut-xsec1,Argoneut-xsec2}, NuTeV\,\cite{NuTeV-xsec}, and CDHS\,\cite{CDHS-xsec} measurements. Not shown are the Gargamelle\,\cite{Gargamelle} and SciBooNE\,\cite{sciboone-inclusive} results, which reported less precise neutrino cross sections within the plotted energy range. This analysis and the IHEP-JINR results are the only measurements of the energy dependence of the charged-current inclusive antineutrino cross section below 5\,GeV. 

In summary, the \lownu flux technique described in this paper yields a total charged-current inclusive cross section shape measurement on CH for both neutrinos and antineutrinos in the 2 through 9~GeV region, which is precisely the region that the current and future oscillation experiments using totally active detectors will occur.  Unlike the measurements of the individual processes (quasi-elastic, pion production) the total cross section measurements agree with the \GENIE simulation and prior data to within their uncertainties but extend to lower neutrino energies than other precision measurements. A byproduct of this analysis is the $\nu_\mu$ and $\bar\nu_\mu$ flux measurements, both for the forward and reverse horn polarities that have been used in the Low Energy NuMI beamline configuration. This is the first time a \lownu-based technique has been used in the NuMI antineutrino-enhanced beam tune. This measurement is the lowest energy application of the \lownu flux technique, and demonstrates that the technique is applicable to future neutrino beams operating at multi-GeV energies. 

\vspace{1mm}
\begin{acknowledgments}
\vspace{-1mm}
\input{Ack} 
\end{acknowledgments}
\bibliographystyle{apsrev4-1}
\bibliography{lownu} 

\clearpage

\input{suppl} 

\begin{turnpage}
\input{covar} 
\end{turnpage}

\end{document}

%% file: authors.tex
\newcommand{\Rutgers}{Rutgers, The State University of New Jersey, Piscataway, New Jersey 08854, USA}
\newcommand{\Hampton}{Hampton University, Dept. of Physics, Hampton, VA 23668, USA}
\newcommand{\Dortmund}{Institute of Physics, Dortmund University, 44221, Germany }
\newcommand{\Otterbein}{Department of Physics, Otterbein University, 1 South Grove Street, Westerville, OH, 43081 USA}
\newcommand{\JMU}{James Madison University, Harrisonburg, Virginia 22807, USA}
\newcommand{\Florida}{University of Florida, Department of Physics, Gainesville, FL 32611}
\newcommand{\UCIrvine}{Department of Physics and Astronomy, University of California, Irvine, Irvine, California 92697-4575, USA}
\newcommand{\CBPF}{Centro Brasileiro de Pesquisas F\'{i}sicas, Rua Dr. Xavier Sigaud 150, Urca, Rio de Janeiro, Rio de Janeiro, 22290-180, Brazil}
\newcommand{\PUCP}{Secci\'{o}n F\'{i}sica, Departamento de Ciencias, Pontificia Universidad Cat\'{o}lica del Per\'{u}, Apartado 1761, Lima, Per\'{u}}
\newcommand{\INRM}{Institute for Nuclear Research of the Russian Academy of Sciences, 117312 Moscow, Russia}
\newcommand{\Jlab}{Jefferson Lab, 12000 Jefferson Avenue, Newport News, VA 23606, USA}
\newcommand{\Pittsburgh}{Department of Physics and Astronomy, University of Pittsburgh, Pittsburgh, Pennsylvania 15260, USA}
\newcommand{\Guanajuato}{Campus Le\'{o}n y Campus Guanajuato, Universidad de Guanajuato, Lascurain de Retana No. 5, Colonia Centro, Guanajuato 36000, Guanajuato M\'{e}xico.}
\newcommand{\Athens}{Department of Physics, University of Athens, GR-15771 Athens, Greece}
\newcommand{\Tufts}{Physics Department, Tufts University, Medford, Massachusetts 02155, USA}
\newcommand{\WM}{Department of Physics, College of William \& Mary, Williamsburg, Virginia 23187, USA}
\newcommand{\FNAL}{Fermi National Accelerator Laboratory, Batavia, Illinois 60510, USA}
\newcommand{\Purdue}{Department of Chemistry and Physics, Purdue University Calumet, Hammond, Indiana 46323, USA}
\newcommand{\MCLA}{Massachusetts College of Liberal Arts, 375 Church Street, North Adams, MA 01247}
\newcommand{\UMD}{Department of Physics, University of Minnesota -- Duluth, Duluth, Minnesota 55812, USA}
\newcommand{\Northwestern}{Northwestern University, Evanston, Illinois 60208}
\newcommand{\UNI}{Universidad Nacional de Ingenier\'{i}a, Apartado 31139, Lima, Per\'{u}}
\newcommand{\Rochester}{University of Rochester, Rochester, New York 14627 USA}
\newcommand{\Austin}{Department of Physics, University of Texas, 1 University Station, Austin, Texas 78712, USA}
\newcommand{\USM}{Departamento de F\'{i}sica, Universidad T\'{e}cnica Federico Santa Mar\'{i}a, Avenida Espa\~{n}a 1680 Casilla 110-V, Valpara\'{i}so, Chile}
\newcommand{\Geneva}{University of Geneva, 1211 Geneva 4, Switzerland}
\newcommand{\Chicago}{Enrico Fermi Institute, University of Chicago, Chicago, IL 60637 USA}
\newcommand{\hired}{}
\newcommand{\OregonState}{Department of Physics, Oregon State University, Corvallis, Oregon 97331, USA}
\newcommand{\oxford}{}
\newcommand{\umiss}{University of Mississippi, University, Mississippi 38677, USA}
\newcommand{\bmeThanks}{now at SLAC National Accelerator Laboratory, Stanford, CA 94309, USA}
\newcommand{\higueraThanks}{now at University of Houston, Houston, TX 77204, USA}
\newcommand{\damartinezThanks}{now at Illinois Institute of Technology, Chicago, IL 60616, USA}
\newcommand{\mcgivernThanks}{now at Fermi National Accelerator Laboratory, Batavia, IL 60510, USA}
\newcommand{\joelmousseauThanks}{now at University of Michigan, Ann Arbor, MI 48109, USA}
\newcommand{\jwolcottThanks}{now at Tufts University, Medford, MA 02155, USA}
\newcommand{\mfcThanks}{now at Oregon State University, Corvallis, OR, 97331, USA}



\author{J.~Devan}                         \affiliation{\WM}
\author{L.~Ren}                           \affiliation{\Pittsburgh}

\author{L.~Aliaga}                        \affiliation{\WM}
\author{O.~Altinok}                       \affiliation{\Tufts}
\author{L.~Bellantoni}                    \affiliation{\FNAL}
\author{A.~Bercellie}                     \affiliation{\Rochester}
\author{M.~Betancourt}                    \affiliation{\FNAL}
\author{A.~Bodek}                         \affiliation{\Rochester}
\author{H.~Budd}                          \affiliation{\Rochester}
\author{T.~Cai}                           \affiliation{\Rochester}
\author{M.F.~Carneiro}\thanks{\mfcThanks} \affiliation{\CBPF} 
\author{H.~da~Motta}                      \affiliation{\CBPF}
\author{S.A.~Dytman}                      \affiliation{\Pittsburgh}
\author{G.A.~D\'{i}az~}                   \affiliation{\Rochester}  \affiliation{\PUCP}
\author{B.~Eberly}\thanks{\bmeThanks}     \affiliation{\Pittsburgh}
\author{E.~Endress}                       \affiliation{\PUCP}
\author{J.~Felix}                         \affiliation{\Guanajuato}
\author{L.~Fields}                        \affiliation{\FNAL}  \affiliation{\Northwestern}
\author{R.~Fine}                          \affiliation{\Rochester}
\author{A.M.~Gago}                        \affiliation{\PUCP}
\author{R.Galindo}                        \affiliation{\USM}
\author{H.~Gallagher}                     \affiliation{\Tufts}
\author{A.~Ghosh}                         \affiliation{\USM}  \affiliation{\CBPF}
\author{R.~Gran}                          \affiliation{\UMD}
\author{D.A.~Harris}                      \affiliation{\FNAL}
\author{A.~Higuera}\thanks{\higueraThanks}  \affiliation{\Rochester}  \affiliation{\Guanajuato}
\author{K.~Hurtado}                       \affiliation{\CBPF}  \affiliation{\UNI}
\author{J.~Kleykamp}                      \affiliation{\Rochester}
\author{M.~Kordosky}                      \affiliation{\WM}
\author{T.~Le}                            \affiliation{\Tufts}  \affiliation{\Rutgers}
\author{E.~Maher}                         \affiliation{\MCLA}
\author{S.~Manly}                         \affiliation{\Rochester}
\author{W.A.~Mann}                        \affiliation{\Tufts}
\author{C.M.~Marshall}                    \affiliation{\Rochester}
\author{D.A.~Martinez~Caicedo}\thanks{\damartinezThanks}  \affiliation{\CBPF}
\author{K.S.~McFarland}                   \affiliation{\Rochester}  \affiliation{\FNAL}
\author{C.L.~McGivern}\thanks{\mcgivernThanks}  \affiliation{\Pittsburgh}
\author{A.M.~McGowan}                     \affiliation{\Rochester}
\author{B.~Messerly}                      \affiliation{\Pittsburgh}
\author{J.~Miller}                        \affiliation{\USM}
\author{A.~Mislivec}                      \affiliation{\Rochester}
\author{J.G.~Morf\'{i}n}                  \affiliation{\FNAL}
\author{J.~Mousseau}\thanks{\joelmousseauThanks}  \affiliation{\Florida}
\author{D.~Naples}                        \affiliation{\Pittsburgh}
\author{J.K.~Nelson}                      \affiliation{\WM}
\author{A.~Norrick}                       \affiliation{\WM}
\author{Nuruzzaman}                       \affiliation{\Rutgers}  \affiliation{\USM}
\author{V.~Paolone}                       \affiliation{\Pittsburgh}
\author{J.~Park}                          \affiliation{\Rochester}
\author{C.E.~Patrick}                     \affiliation{\Northwestern}
\author{G.N.~Perdue}                      \affiliation{\FNAL}  \affiliation{\Rochester}
\author{M.A.~Ramirez}                     \affiliation{\Guanajuato}
\author{R.D.~Ransome}                     \affiliation{\Rutgers}
\author{H.~Ray}                           \affiliation{\Florida}
\author{D.~Rimal}                         \affiliation{\Florida}
\author{P.A.~Rodrigues}                   \affiliation{\Rochester} \affiliation{\umiss}
\author{D.~Ruterbories}                   \affiliation{\Rochester}
\author{H.~Schellman}                     \affiliation{\OregonState}  \affiliation{\Northwestern}
\author{C.J.~Solano~Salinas}              \affiliation{\UNI}
\author{B.G.~Tice}                        \affiliation{\Rutgers}
\author{E.~Valencia}                      \affiliation{\WM}  \affiliation{\Guanajuato}
\author{J.~Wolcott}\thanks{\jwolcottThanks}  \affiliation{\Rochester}
\author{M.Wospakrik}                      \affiliation{\Florida}

\collaboration{The MINERvA  Collaboration}\ \noaffiliation
\date{\today}

\noaffiliation

%% file: Ack.tex
This work was supported by the Fermi National Accelerator Laboratory under US Department of Energy contract No. DE-AC02-07CH11359 which included the MINERvA construction project. Construction support was also granted by the United States National Science Foundation under Award PHY-0619727 and by the University of Rochester. Support for participating scientists was provided by NSF and DOE (USA), by CAPES and CNPq (Brazil), by CoNaCyT (Mexico), by CONICYT (Chile), by CONCYTEC, DGI-PUCP and IDI/IGI-UNI (Peru), and by the Latin American Center for Physics (CLAF). We thank the MINOS Collaboration for use of its near detector data. We acknowledge the dedicated work of the Fermilab staff responsible for the operation and maintenance of the beamline and detector and the Fermilab Computing Division for support of data processing.

%% file: suppl.tex
\onecolumngrid

\vspace{1in}
\centerline{\textbf{\Large Supplemental materials}}

\begin{turnpage}
\begin{table}[pht]
   \setlength{\tabcolsep}{4pt}
    \begin{tabular}{ccccc|cccc|cccccc}
    \hline
 Bin & $\Phi(E)$ & 	 Stat.  & 	 Norm. & 	 Shape. & 	 Muon  & 	 Recoil  & 	 Inter.  & 	 Other & Inter. &  	 Inter. &  	 Nuclear &  	 Nuclear &  	 Non- &  	 Inter. \\
 (GeV) 	    &                  & 	Uncert.  & 	 Uncert. & 	 Uncert. 	& 	  Reco. & 	 Reco.  & 	  Model & 	 & MaRES &  Rvn1pi &  	 RPA &  	 RPA+MEC &  	 Reweightable & Other \\
    \hline
 2--3 & 	 76.046 &  	 0.927 &  	 2.801 &  	 6.891 &  	 5.705 &  	 1.585 &  	 3.262 &  	 1.445 &  	 0.779 &  	 0.217 &  	 2.630 &  	 1.427 &  	 0.573 &  	 0.819 \\ 
 3--4 & 	 77.528 &  	 0.838 &  	 2.855 &  	 4.229 &  	 2.732 &  	 1.897 &  	 2.176 &  	 1.482 &  	 0.612 &  	 0.078 &  	 1.689 &  	 0.366 &  	 0.585 &  	 0.867 \\ 
 4--5 & 	 29.776 &  	 0.490 &  	 1.097 &  	 1.981 &  	 1.565 &  	 0.835 &  	 0.672 &  	 0.592 &  	 0.204 &  	 0.018 &  	 0.450 &  	 0.062 &  	 0.225 &  	 0.309 \\ 
 5--6 & 	 12.243 &  	 0.180 &  	 0.451 &  	 0.621 &  	 0.420 &  	 0.249 &  	 0.295 &  	 0.245 &  	 0.106 &  	 0.044 &  	 0.017 &  	 0.214 &  	 0.106 &  	 0.108 \\ 
 6--7 & 	 7.908 &  	 0.151 &  	 0.291 &  	 0.343 &  	 0.146 &  	 0.090 &  	 0.234 &  	 0.183 &  	 0.095 &  	 0.056 &  	 0.022 &  	 0.166 &  	 0.068 &  	 0.085 \\ 
 7--9 & 	 10.435 &  	 0.181 &  	 0.384 &  	 0.412 &  	 0.174 &  	 0.115 &  	 0.302 &  	 0.187 &  	 0.143 &  	 0.086 &  	 0.025 &  	 0.207 &  	 0.090 &  	 0.107 \\ 
 9--12 & 	 9.780 &  	 0.123 &  	 0.360 &  	 0.348 &  	 0.165 &  	 0.118 &  	 0.240 &  	 0.148 &  	 0.133 &  	 0.085 &  	 0.005 &  	 0.144 &  	 0.051 &  	 0.096 \\ 
 12--15 & 	 6.115 &  	 0.093 &  	 0.225 &  	 0.231 &  	 0.116 &  	 0.083 &  	 0.153 &  	 0.098 &  	 0.067 &  	 0.054 &  	 0.007 &  	 0.106 &  	 0.012 &  	 0.065 \\ 
 15--18 & 	 3.477 &  	 0.064 &  	 0.128 &  	 0.136 &  	 0.080 &  	 0.041 &  	 0.085 &  	 0.057 &  	 0.043 &  	 0.029 &  	 0.007 &  	 0.053 &  	 0.007 &  	 0.040 \\ 
 18--22 & 	 2.620 &  	 0.056 &  	 0.096 &  	 0.130 &  	 0.094 &  	 0.040 &  	 0.064 &  	 0.050 &  	 0.038 &  	 0.025 &  	 0.004 &  	 0.035 &  	 0.005 &  	 0.026 \\ 
 22--26 & 	 1.331 &  	 0.039 &  	 0.049 &  	 0.076 &  	 0.060 &  	 0.021 &  	 0.031 &  	 0.028 &  	 0.018 &  	 0.012 &  	 0.003 &  	 0.016 &  	 0.003 &  	 0.015 \\ 
 26--30 & 	 0.751 &  	 0.028 &  	 0.028 &  	 0.044 &  	 0.032 &  	 0.012 &  	 0.017 &  	 0.021 &  	 0.009 &  	 0.007 &  	 0.002 &  	 0.009 &  	 0.001 &  	 0.008 \\ 
 30--36 & 	 0.726 &  	 0.028 &  	 0.027 &  	 0.034 &  	 0.020 &  	 0.012 &  	 0.018 &  	 0.016 &  	 0.012 &  	 0.007 &  	 0.002 &  	 0.008 &  	 0.001 &  	 0.008 \\ 
 36--42 & 	 0.497 &  	 0.025 &  	 0.018 &  	 0.024 &  	 0.012 &  	 0.011 &  	 0.012 &  	 0.013 &  	 0.005 &  	 0.006 &  	 0.000 &  	 0.007 &  	 0.001 &  	 0.006 \\ 
 42--50 & 	 0.452 &  	 0.025 &  	 0.017 &  	 0.026 &  	 0.013 &  	 0.012 &  	 0.013 &  	 0.014 &  	 0.006 &  	 0.005 &  	 0.001 &  	 0.008 &  	 0.001 &  	 0.006 \\
 \hline
 \end{tabular}
\caption{Neutrino flux in the forward horn current (FHC) beam tune (in units of $\nu$/m$^2$/$10^6$\,POT) and uncertainties including statistical, external normalization, and shape-dependent systematic errors. The shape-dependent uncertainties are further broken down into the largest categories including reconstruction uncertainties, uncertainties associated with the nuclear model, and the remaining categories of uncertainties outlined in the text. The nuclear model uncertainties are further broken down into their largest components. Note that the values and uncertainties are not bin width normalized.} 
\label{tbl:FHC-nu-flux}
\end{table}

\begin{table}[hpt]
   \setlength{\tabcolsep}{4pt}
    \begin{tabular}{ccccc|cccc|cccccc}
    \hline
 Bin & $\Phi(E)$ & 	 Stat.  & 	 Norm. & 	 Shape. & 	 Muon  & 	 Recoil  & 	 Inter.  & 	 Other & Inter. &  	 Inter. &  	 Nuclear &  	 Nuclear &  	 Non- &  	 Inter. \\
 (GeV) 	    &                  & 	Uncert.  & 	 Uncert. & 	 Uncert. 	& 	  Reco. & 	 Reco.  & 	  Model & 	 & MaRES &  Rvn1pi &  	 RPA &  	 RPA+MEC &  	 Reweightable & Other \\
    \hline
  2--3 &  	 3.344 &  	 0.157 &  	 0.351 &  	 0.392 &  	 0.205 &  	 0.030 &  	 0.310 &  	 0.126 &  	 0.003 &  	 0.021 &  	 0.238 &  	 0.173 &  	 0.025 &  	 0.092 \\ 
  3--4 &  	 3.496 &  	 0.143 &  	 0.367 &  	 0.303 &  	 0.186 &  	 0.039 &  	 0.208 &  	 0.115 &  	 0.002 &  	 0.013 &  	 0.166 &  	 0.114 &  	 0.026 &  	 0.038 \\ 
  4--5 &  	 3.161 &  	 0.130 &  	 0.332 &  	 0.211 &  	 0.125 &  	 0.055 &  	 0.135 &  	 0.090 &  	 0.004 &  	 0.013 &  	 0.109 &  	 0.070 &  	 0.024 &  	 0.026 \\ 
  5--6 &  	 2.817 &  	 0.086 &  	 0.296 &  	 0.144 &  	 0.093 &  	 0.035 &  	 0.064 &  	 0.082 &  	 0.012 &  	 0.007 &  	 0.051 &  	 0.021 &  	 0.024 &  	 0.015 \\ 
  6--7 &  	 2.497 &  	 0.079 &  	 0.262 &  	 0.120 &  	 0.086 &  	 0.025 &  	 0.044 &  	 0.068 &  	 0.003 &  	 0.006 &  	 0.034 &  	 0.009 &  	 0.022 &  	 0.012 \\ 
  7--9 &  	 4.256 &  	 0.102 &  	 0.447 &  	 0.154 &  	 0.103 &  	 0.036 &  	 0.057 &  	 0.093 &  	 0.000 &  	 0.013 &  	 0.035 &  	 0.006 &  	 0.037 &  	 0.020 \\ 
  9--12 &  	 4.017 &  	 0.075 &  	 0.422 &  	 0.104 &  	 0.063 &  	 0.009 &  	 0.048 &  	 0.066 &  	 0.013 &  	 0.010 &  	 0.004 &  	 0.038 &  	 0.021 &  	 0.012 \\ 
  12--15 &  	 2.287 &  	 0.056 &  	 0.240 &  	 0.074 &  	 0.049 &  	 0.013 &  	 0.034 &  	 0.042 &  	 0.019 &  	 0.007 &  	 0.001 &  	 0.023 &  	 0.005 &  	 0.014 \\ 
  15--18 &  	 1.217 &  	 0.040 &  	 0.128 &  	 0.049 &  	 0.036 &  	 0.011 &  	 0.020 &  	 0.025 &  	 0.010 &  	 0.006 &  	 0.001 &  	 0.012 &  	 0.002 &  	 0.009 \\ 
  18--22 &  	 0.838 &  	 0.032 &  	 0.088 &  	 0.037 &  	 0.026 &  	 0.011 &  	 0.014 &  	 0.020 &  	 0.004 &  	 0.008 &  	 0.000 &  	 0.006 &  	 0.002 &  	 0.008 \\ 
  22--26 &  	 0.380 &  	 0.021 &  	 0.040 &  	 0.027 &  	 0.022 &  	 0.005 &  	 0.007 &  	 0.013 &  	 0.002 &  	 0.004 &  	 0.000 &  	 0.003 &  	 0.001 &  	 0.004 \\ 
  26--30 &  	 0.179 &  	 0.014 &  	 0.019 &  	 0.014 &  	 0.011 &  	 0.004 &  	 0.003 &  	 0.007 &  	 0.001 &  	 0.002 &  	 0.000 &  	 0.001 &  	 0.000 &  	 0.002 \\ 
  30--36 &  	 0.145 &  	 0.013 &  	 0.015 &  	 0.010 &  	 0.007 &  	 0.004 &  	 0.003 &  	 0.006 &  	 0.001 &  	 0.001 &  	 0.000 &  	 0.001 &  	 0.000 &  	 0.002 \\ 
  36--42 &  	 0.081 &  	 0.010 &  	 0.008 &  	 0.006 &  	 0.003 &  	 0.002 &  	 0.001 &  	 0.004 &  	 0.000 &  	 0.000 &  	 0.000 &  	 0.000 &  	 0.000 &  	 0.001 \\ 
  42--50 &  	 0.054 &  	 0.008 &  	 0.006 &  	 0.005 &  	 0.003 &  	 0.002 &  	 0.001 &  	 0.003 &  	 0.000 &  	 0.000 &  	 0.000 &  	 0.000 &  	 0.000 &  	 0.001 \\ 
  \hline
\end{tabular}
\caption{Antineutrino flux in the forward horn current (FHC) beam tune (in units of $\nu$/m$^2$/$10^6$\,POT) and uncertainties including statistical, external normalization, and shape-dependent systematic errors. The shape-dependent uncertainties are further broken down into the largest categories including reconstruction uncertainties, uncertainties associated with the nuclear model, and the remaining categories of uncertainties outlined in the text. The nuclear model uncertainties are further broken down into their largest components. Note that the values and uncertainties are not bin width normalized.}
\label{tbl:FHC-anti-flux}
\end{table}

\begin{table}[hpt]
   \setlength{\tabcolsep}{4pt}
    \begin{tabular}{ccccc|cccc|cccccc}
    \hline
 Bin & $\Phi(E)$ & 	 Stat.  & 	 Norm. & 	 Shape. & 	 Muon  & 	 Recoil  & 	 Inter.  & 	 Other & Inter. &  	 Inter. &  	 Nuclear &  	 Nuclear &  	 Non- &  	 Inter. \\
 (GeV) 	    &                  & 	Uncert.  & 	 Uncert. & 	 Uncert. 	& 	  Reco. & 	 Reco.  & 	  Model & 	 & MaRES &  Rvn1pi &  	 RPA &  	 RPA+MEC &  	 Reweightable & Other \\
    \hline
  2--3 &  	 69.546 &  	 1.357 &  	 7.492 &  	 8.885 &  	 4.982 &  	 1.777 &  	 6.926 &  	 1.768 &  	 0.392 &  	 0.089 &  	 4.944 &  	 4.697 &  	 0.524 &  	 1.009 \\ 
  3--4 &  	 64.995 &  	 1.148 &  	 7.002 &  	 5.556 &  	 1.763 &  	 1.934 &  	 4.627 &  	 1.633 &  	 0.394 &  	 0.068 &  	 3.269 &  	 3.035 &  	 0.490 &  	 1.035 \\ 
  4--5 &  	 23.078 &  	 0.634 &  	 2.486 &  	 2.234 &  	 1.239 &  	 0.896 &  	 1.515 &  	 0.608 &  	 0.159 &  	 0.026 &  	 1.057 &  	 0.985 &  	 0.174 &  	 0.385 \\ 
  5--6 &  	 7.646 &  	 0.243 &  	 0.824 &  	 0.481 &  	 0.276 &  	 0.285 &  	 0.170 &  	 0.212 &  	 0.054 &  	 0.025 &  	 0.079 &  	 0.050 &  	 0.066 &  	 0.104 \\ 
  6--7 &  	 4.902 &  	 0.198 &  	 0.528 &  	 0.237 &  	 0.107 &  	 0.130 &  	 0.073 &  	 0.151 &  	 0.014 &  	 0.012 &  	 0.034 &  	 0.018 &  	 0.042 &  	 0.042 \\ 
  7--9 &  	 6.012 &  	 0.226 &  	 0.648 &  	 0.239 &  	 0.108 &  	 0.110 &  	 0.086 &  	 0.162 &  	 0.015 &  	 0.012 &  	 0.004 &  	 0.017 &  	 0.052 &  	 0.062 \\ 
  9--12 &  	 5.101 &  	 0.155 &  	 0.550 &  	 0.123 &  	 0.072 &  	 0.021 &  	 0.032 &  	 0.092 &  	 0.005 &  	 0.000 &  	 0.001 &  	 0.010 &  	 0.027 &  	 0.012 \\ 
  12--15 &  	 2.576 &  	 0.100 &  	 0.278 &  	 0.071 &  	 0.038 &  	 0.019 &  	 0.025 &  	 0.051 &  	 0.002 &  	 0.004 &  	 0.006 &  	 0.004 &  	 0.005 &  	 0.022 \\ 
  15--18 &  	 1.400 &  	 0.069 &  	 0.151 &  	 0.051 &  	 0.037 &  	 0.003 &  	 0.015 &  	 0.032 &  	 0.005 &  	 0.002 &  	 0.004 &  	 0.003 &  	 0.003 &  	 0.013 \\ 
  18--22 &  	 0.909 &  	 0.054 &  	 0.098 &  	 0.049 &  	 0.041 &  	 0.007 &  	 0.010 &  	 0.024 &  	 0.003 &  	 0.001 &  	 0.002 &  	 0.001 &  	 0.002 &  	 0.008 \\ 
  22--26 &  	 0.399 &  	 0.035 &  	 0.043 &  	 0.027 &  	 0.022 &  	 0.004 &  	 0.005 &  	 0.015 &  	 0.001 &  	 0.003 &  	 0.000 &  	 0.001 &  	 0.001 &  	 0.004 \\ 
  26--30 &  	 0.193 &  	 0.025 &  	 0.021 &  	 0.015 &  	 0.011 &  	 0.002 &  	 0.004 &  	 0.009 &  	 0.001 &  	 0.002 &  	 0.000 &  	 0.001 &  	 0.000 &  	 0.004 \\ 
  30--36 &  	 0.178 &  	 0.022 &  	 0.019 &  	 0.013 &  	 0.005 &  	 0.005 &  	 0.005 &  	 0.009 &  	 0.000 &  	 0.001 &  	 0.001 &  	 0.001 &  	 0.000 &  	 0.003 \\ 
  36--42 &  	 0.110 &  	 0.020 &  	 0.012 &  	 0.011 &  	 0.006 &  	 0.003 &  	 0.002 &  	 0.008 &  	 0.000 &  	 0.001 &  	 0.001 &  	 0.001 &  	 0.000 &  	 0.001 \\ 
  42--50 &  	 0.062 &  	 0.015 &  	 0.007 &  	 0.007 &  	 0.004 &  	 0.001 &  	 0.002 &  	 0.005 &  	 0.001 &  	 0.001 &  	 0.000 &  	 0.000 &  	 0.000 &  	 0.001 \\     
   \hline
\end{tabular}
\caption{Antineutrino flux in the reverse horn current (RHC) beam tune (in units of $\nu$/m$^2$/$10^6$\,POT) and uncertainties including statistical, external normalization, and shape-dependent systematic errors. The shape-dependent uncertainties are further broken down into the largest categories including reconstruction uncertainties, uncertainties associated with the nuclear model, and the remaining categories of uncertainties outlined in the text. The nuclear model uncertainties are further broken down into their largest components. Note that the values and uncertainties are not bin width normalized.}
\label{tbl:RHC-anti-flux}
\end{table}

\begin{table}[hpt]
   \setlength{\tabcolsep}{4pt}
    \begin{tabular}{ccccc|cccc|cccccc}
    \hline
 Bin & $\Phi(E)$ & 	 Stat.  & 	 Norm. & 	 Shape. & 	 Muon  & 	 Recoil  & 	 Inter.  & 	 Other & Inter. &  	 Inter. &  	 Nuclear &  	 Nuclear &  	 Non- &  	 Inter. \\
 (GeV) 	    &                  & 	Uncert.  & 	 Uncert. & 	 Uncert. 	& 	  Reco. & 	 Reco.  & 	  Model & 	 & MaRES &  Rvn1pi &  	 RPA &  	 RPA+MEC &  	 Reweightable & Other \\
    \hline
 2--3 & 	 3.995 &  	 0.305 &  	 0.147 &  	 0.300 &  	 0.216 &  	 0.048 &  	 0.130 &  	 0.156 &  	 0.089 &  	 0.005 &  	 0.052 &  	 0.016 &  	 0.030 &  	 0.070 \\ 
 3--4 & 	 4.263 &  	 0.290 &  	 0.157 &  	 0.292 &  	 0.193 &  	 0.063 &  	 0.145 &  	 0.152 &  	 0.084 &  	 0.021 &  	 0.046 &  	 0.047 &  	 0.032 &  	 0.089 \\ 
 4--5 & 	 3.806 &  	 0.271 &  	 0.140 &  	 0.256 &  	 0.148 &  	 0.067 &  	 0.141 &  	 0.138 &  	 0.074 &  	 0.038 &  	 0.049 &  	 0.048 &  	 0.029 &  	 0.083 \\ 
 5--6 & 	 3.701 &  	 0.153 &  	 0.136 &  	 0.229 &  	 0.150 &  	 0.047 &  	 0.118 &  	 0.116 &  	 0.066 &  	 0.029 &  	 0.049 &  	 0.048 &  	 0.032 &  	 0.052 \\ 
 6--7 & 	 3.864 &  	 0.155 &  	 0.142 &  	 0.241 &  	 0.155 &  	 0.069 &  	 0.131 &  	 0.111 &  	 0.073 &  	 0.043 &  	 0.035 &  	 0.064 &  	 0.033 &  	 0.057 \\ 
 7--9 & 	 7.265 &  	 0.218 &  	 0.268 &  	 0.407 &  	 0.210 &  	 0.149 &  	 0.266 &  	 0.169 &  	 0.137 &  	 0.100 &  	 0.038 &  	 0.154 &  	 0.063 &  	 0.107 \\ 
 9--12 & 	 8.221 &  	 0.167 &  	 0.303 &  	 0.393 &  	 0.143 &  	 0.166 &  	 0.298 &  	 0.133 &  	 0.158 &  	 0.095 &  	 0.012 &  	 0.202 &  	 0.043 &  	 0.110 \\ 
 12--15 & 	 4.920 &  	 0.130 &  	 0.181 &  	 0.245 &  	 0.083 &  	 0.093 &  	 0.192 &  	 0.087 &  	 0.103 &  	 0.062 &  	 0.003 &  	 0.127 &  	 0.010 &  	 0.078 \\ 
 15--18 & 	 2.960 &  	 0.102 &  	 0.109 &  	 0.167 &  	 0.077 &  	 0.059 &  	 0.122 &  	 0.060 &  	 0.066 &  	 0.040 &  	 0.007 &  	 0.082 &  	 0.006 &  	 0.044 \\ 
 18--22 & 	 2.203 &  	 0.087 &  	 0.081 &  	 0.138 &  	 0.066 &  	 0.050 &  	 0.098 &  	 0.052 &  	 0.057 &  	 0.033 &  	 0.006 &  	 0.064 &  	 0.004 &  	 0.034 \\ 
 22--26 & 	 1.110 &  	 0.059 &  	 0.041 &  	 0.074 &  	 0.037 &  	 0.027 &  	 0.050 &  	 0.030 &  	 0.026 &  	 0.021 &  	 0.005 &  	 0.032 &  	 0.002 &  	 0.018 \\ 
 26--30 & 	 0.579 &  	 0.041 &  	 0.021 &  	 0.045 &  	 0.027 &  	 0.018 &  	 0.022 &  	 0.022 &  	 0.011 &  	 0.008 &  	 0.003 &  	 0.015 &  	 0.001 &  	 0.008 \\ 
 30--36 & 	 0.542 &  	 0.041 &  	 0.020 &  	 0.037 &  	 0.022 &  	 0.019 &  	 0.017 &  	 0.017 &  	 0.006 &  	 0.006 &  	 0.001 &  	 0.012 &  	 0.001 &  	 0.008 \\ 
 36--42 & 	 0.371 &  	 0.033 &  	 0.014 &  	 0.024 &  	 0.006 &  	 0.011 &  	 0.016 &  	 0.014 &  	 0.010 &  	 0.005 &  	 0.000 &  	 0.008 &  	 0.001 &  	 0.007 \\ 
 42--50 & 	 0.394 &  	 0.039 &  	 0.015 &  	 0.031 &  	 0.007 &  	 0.010 &  	 0.023 &  	 0.017 &  	 0.015 &  	 0.005 &  	 0.003 &  	 0.012 &  	 0.001 &  	 0.010 \\ 
  \hline
 \end{tabular}
\caption{Neutrino flux in the reverse horn current (RHC) beam tune (in units of $\nu$/m$^2$/$10^6$\,POT) and uncertainties including statistical, external normalization, and shape-dependent systematic errors. The shape-dependent uncertainties are further broken down into the largest categories including reconstruction uncertainties, uncertainties associated with the nuclear model, and the remaining categories of uncertainties outlined in the text. The nuclear model uncertainties are further broken down into their largest components. Note that the values and uncertainties are not bin width normalized.}
    \label{tbl:RHC-nu-flux}
\end{table}

\begin{table}[hpt]
   \setlength{\tabcolsep}{3pt}
    \begin{tabular}{cccccc|cccc|cccccc}
    \hline
 Bin & $<E>$  & $\sigma(E_\nu)$ & 	 Stat.  & 	 Norm. & 	 Shape. & 	 Muon  & 	 Recoil  & 	 Inter.  & 	 Other & Inter. &  	 Inter. &  	 Nuclear &  	 Nuclear &  	 Non- &  	 Inter. \\
 (GeV) 	    &  (GeV) &                & 	Uncert.  & 	 Uncert. & 	 Uncert. 	& 	  Reco. & 	 Reco.  & 	  Model & 	 & MaRES &  Rvn1pi &  	 RPA &  	 RPA+MEC &  	 Reweightable & Other \\
 \hline
 2--3 & 	 2.63 & 	 0.767 &  	 0.007 &  	 0.028 &  	 0.106 &  	 0.005 &  	 0.072 &  	 0.077 &  	 0.009 &  	 0.044 &  	 0.029 &  	 0.030 &  	 0.015 &  	 0.005 &  	 0.008 \\ 
 3--4 & 	 3.51 & 	 0.730 &  	 0.007 &  	 0.027 &  	 0.053 &  	 0.004 &  	 0.034 &  	 0.040 &  	 0.009 &  	 0.014 &  	 0.019 &  	 0.006 &  	 0.019 &  	 0.005 &  	 0.003 \\ 
 4--5 & 	 4.44 & 	 0.751 &  	 0.011 &  	 0.028 &  	 0.039 &  	 0.015 &  	 0.013 &  	 0.032 &  	 0.010 &  	 0.003 &  	 0.009 &  	 0.004 &  	 0.027 &  	 0.005 &  	 0.001 \\ 
 5--6 & 	 5.47 & 	 0.726 &  	 0.008 &  	 0.027 &  	 0.024 &  	 0.010 &  	 0.007 &  	 0.020 &  	 0.008 &  	 0.009 &  	 0.000 &  	 0.000 &  	 0.015 &  	 0.006 &  	 0.000 \\ 
 6--7 & 	 6.49 & 	 0.704 &  	 0.011 &  	 0.026 &  	 0.018 &  	 0.006 &  	 0.006 &  	 0.013 &  	 0.007 &  	 0.006 &  	 0.000 &  	 0.002 &  	 0.007 &  	 0.006 &  	 0.000 \\ 
 7--9 & 	 7.97 & 	 0.720 &  	 0.011 &  	 0.027 &  	 0.016 &  	 0.006 &  	 0.009 &  	 0.009 &  	 0.007 &  	 0.004 &  	 0.000 &  	 0.002 &  	 0.002 &  	 0.006 &  	 0.000 \\ 
 9--12 & 	 10.45 & 	 0.706 &  	 0.007 &  	 0.026 &  	 0.003 &  	 0.000 &  	 0.000 &  	 0.002 &  	 0.003 &  	 0.000 &  	 0.001 &  	 0.000 &  	 0.000 &  	 0.000 &  	 0.000 \\ 
 12--15 & 	 13.45 & 	 0.677 &  	 0.008 &  	 0.025 &  	 0.014 &  	 0.004 &  	 0.009 &  	 0.010 &  	 0.004 &  	 0.001 &  	 0.002 &  	 0.001 &  	 0.002 &  	 0.009 &  	 0.000 \\ 
 15--18 & 	 16.43 & 	 0.701 &  	 0.011 &  	 0.026 &  	 0.022 &  	 0.009 &  	 0.016 &  	 0.010 &  	 0.005 &  	 0.001 &  	 0.000 &  	 0.001 &  	 0.001 &  	 0.009 &  	 0.000 \\ 
 18--22 & 	 19.90 & 	 0.721 &  	 0.014 &  	 0.027 &  	 0.030 &  	 0.015 &  	 0.023 &  	 0.011 &  	 0.006 &  	 0.003 &  	 0.002 &  	 0.001 &  	 0.002 &  	 0.009 &  	 0.000 \\ 
 22--26 & 	 23.88 & 	 0.758 &  	 0.020 &  	 0.028 &  	 0.042 &  	 0.019 &  	 0.034 &  	 0.013 &  	 0.009 &  	 0.004 &  	 0.003 &  	 0.001 &  	 0.003 &  	 0.010 &  	 0.001 \\ 
 26--30 & 	 27.88 & 	 0.772 &  	 0.027 &  	 0.028 &  	 0.049 &  	 0.017 &  	 0.042 &  	 0.015 &  	 0.011 &  	 0.004 &  	 0.005 &  	 0.002 &  	 0.004 &  	 0.010 &  	 0.001 \\ 
 30--36 & 	 32.80 & 	 0.710 &  	 0.026 &  	 0.026 &  	 0.038 &  	 0.008 &  	 0.032 &  	 0.015 &  	 0.011 &  	 0.008 &  	 0.004 &  	 0.002 &  	 0.004 &  	 0.009 &  	 0.001 \\ 
 36--42 & 	 38.87 & 	 0.671 &  	 0.032 &  	 0.025 &  	 0.029 &  	 0.005 &  	 0.021 &  	 0.015 &  	 0.014 &  	 0.004 &  	 0.006 &  	 0.000 &  	 0.006 &  	 0.009 &  	 0.001 \\ 
 42--50 & 	 45.72 &  	 0.751 &  	 0.040 &  	 0.028 &  	 0.033 &  	 0.011 &  	 0.015 &  	 0.020 &  	 0.019 &  	 0.007 &  	 0.006 &  	 0.002 &  	 0.010 &  	 0.010 &  	 0.001 \\ 
\hline
\end{tabular}
\caption{Isoscalar-corrected neutrino charged-current inclusive cross section (in units of $10^{-38}$ cm$^2$/GeV/nucleon) and uncertainties including statistical, external normalization, and shape-dependent systematic errors. The shape-dependent uncertainties are further broken down into the largest categories including reconstruction uncertainties, uncertainties associated with the nuclear model, and the remaining categories of uncertainties outlined in the text. The nuclear model uncertainties are further broken down into their largest components. Note that the values and uncertainties are not bin width normalized.} 
\label{tbl:FHC-nu-xsec}
\end{table}

\begin{table}[ptbh]
   \setlength{\tabcolsep}{3pt}
    \begin{tabular}{cccccc|cccc|cccccc}
    \hline
 Bin & $<E>$  & $\sigma(E_\nu)$ & 	 Stat.  & 	 Norm. & 	 Shape. & 	 Muon  & 	 Recoil  & 	 Inter.  & 	 Other & Inter. &  	 Inter. &  	 Nuclear &  	 Nuclear &  	 Non- &  	 Inter. \\
 (GeV) 	    &  (GeV) &                & 	Uncert.  & 	 Uncert. & 	 Uncert. 	& 	  Reco. & 	 Reco.  & 	  Model & 	 & MaRES &  Rvn1pi &  	 RPA &  	 RPA+MEC &  	 Reweightable & Other \\
    \hline
 2--3 & 	 2.61 & 	 0.365 &  	 0.004 &  	 0.039 &  	 0.042 &  	 0.001 &  	 0.035 &  	 0.022 &  	 0.007 &  	 0.015 &  	 0.010 &  	 0.001 &  	 0.001 &  	 0.002 &  	 0.005 \\ 
 3--4 & 	 3.48 & 	 0.333 &  	 0.004 &  	 0.036 &  	 0.024 &  	 0.001 &  	 0.015 &  	 0.017 &  	 0.007 &  	 0.006 &  	 0.006 &  	 0.006 &  	 0.011 &  	 0.002 &  	 0.003 \\ 
 4--5 & 	 4.43 & 	 0.334 &  	 0.008 &  	 0.036 &  	 0.022 &  	 0.004 &  	 0.006 &  	 0.019 &  	 0.007 &  	 0.001 &  	 0.003 &  	 0.010 &  	 0.015 &  	 0.002 &  	 0.002 \\ 
 5--6 & 	 5.46 &  	 0.349 &  	 0.008 &  	 0.038 &  	 0.012 &  	 0.003 &  	 0.002 &  	 0.009 &  	 0.007 &  	 0.002 &  	 0.001 &  	 0.002 &  	 0.006 &  	 0.003 &  	 0.001 \\ 
 6--7 & 	 6.47 & 	 0.331 &  	 0.010 &  	 0.036 &  	 0.011 &  	 0.004 &  	 0.005 &  	 0.006 &  	 0.007 &  	 0.001 &  	 0.000 &  	 0.002 &  	 0.004 &  	 0.003 &  	 0.001 \\ 
 7--9 & 	 7.95 & 	 0.341 &  	 0.010 &  	 0.037 &  	 0.010 &  	 0.002 &  	 0.005 &  	 0.005 &  	 0.007 &  	 0.001 &  	 0.000 &  	 0.000 &  	 0.001 &  	 0.003 &  	 0.001 \\ 
 9--12 & 	 10.41 & 	 0.320 &  	 0.007 &  	 0.034 &  	 0.002 &  	 0.000 &  	 0.000 &  	 0.000 &  	 0.002 &  	 0.000 &  	 0.000 &  	 0.000 &  	 0.000 &  	 0.000 &  	 0.000 \\ 
 12--15 & 	 13.40 &  	 0.344 &  	 0.010 &  	 0.037 &  	 0.008 &  	 0.000 &  	 0.004 &  	 0.006 &  	 0.003 &  	 0.001 &  	 0.001 &  	 0.001 &  	 0.002 &  	 0.004 &  	 0.001 \\ 
 15--18 & 	 16.41 & 	 0.329 &  	 0.013 &  	 0.035 &  	 0.011 &  	 0.002 &  	 0.008 &  	 0.006 &  	 0.004 &  	 0.001 &  	 0.001 &  	 0.001 &  	 0.003 &  	 0.004 &  	 0.001 \\ 
 18--22 & 	 19.82 & 	 0.343 &  	 0.017 &  	 0.037 &  	 0.017 &  	 0.007 &  	 0.013 &  	 0.006 &  	 0.006 &  	 0.001 &  	 0.001 &  	 0.001 &  	 0.002 &  	 0.004 &  	 0.001 \\ 
 22--26 & 	 23.85 & 	 0.375 &  	 0.029 &  	 0.040 &  	 0.023 &  	 0.010 &  	 0.017 &  	 0.007 &  	 0.009 &  	 0.001 &  	 0.001 &  	 0.000 &  	 0.000 &  	 0.005 &  	 0.001 \\ 
 26--30 & 	 27.81 & 	 0.392 &  	 0.047 &  	 0.042 &  	 0.028 &  	 0.006 &  	 0.021 &  	 0.009 &  	 0.014 &  	 0.001 &  	 0.003 &  	 0.001 &  	 0.001 &  	 0.005 &  	 0.003 \\ 
 30--36 & 	 32.71 & 	 0.309 &  	 0.034 &  	 0.033 &  	 0.026 &  	 0.003 &  	 0.020 &  	 0.009 &  	 0.012 &  	 0.001 &  	 0.001 &  	 0.002 &  	 0.002 &  	 0.004 &  	 0.004 \\ 
 36--42 & 	 38.68 & 	 0.286 &  	 0.048 &  	 0.031 &  	 0.028 &  	 0.010 &  	 0.017 &  	 0.010 &  	 0.017 &  	 0.001 &  	 0.002 &  	 0.003 &  	 0.003 &  	 0.004 &  	 0.003 \\ 
 42--50 & 	 45.44 & 	 0.314 &  	 0.068 &  	 0.034 &  	 0.038 &  	 0.009 &  	 0.026 &  	 0.011 &  	 0.024 &  	 0.005 &  	 0.002 &  	 0.002 &  	 0.000 &  	 0.004 &  	 0.006 \\ 
 \hline
 \end{tabular}
\caption{Isoscalar-corrected antineutrino charged-current inclusive cross section (in units of $10^{-38}$ cm$^2$/GeV/nucleon) and uncertainties including statistical, external normalization, and shape-dependent systematic errors. The shape-dependent uncertainties are further broken down into the largest categories including reconstruction uncertainties, uncertainties associated with the nuclear model, and the remaining categories of uncertainties outlined in the text. The nuclear model uncertainties are further broken down into their largest components. Note that the values and uncertainties are not bin width normalized.} 
\label{tbl:RHC-anti-xsec}
\end{table}

\end{turnpage}

%% file: covar.tex
\begin{table}[tb]
    {\renewcommand{\tabcolsep}{0.1cm} \begin{tabular}{c|ccccccccccccccc}
    \hline
           &   2--3 &   3--4 &   4--5 &   5--6 &   6--7 &   7--9 &  9--12 & 12--15 & 15--18 & 18--22 & 22--26 & 26--30 & 30--36 & 36--42 & 42--50 \\
    \hline
    2--3   & 56.077 & 19.744 & -1.529 & -0.807 & -0.005 &  0.345 &  0.632 &  0.507 & -0.124 & -0.309 & -0.213 & -0.109 & -0.045 & -0.032 & -0.052 \\
    3--4   & 19.744 & 26.651 &  4.688 &  1.180 &  0.613 &  0.753 &  0.555 &  0.317 &  0.158 &  0.084 &  0.047 &  0.027 &  0.030 &  0.011 &  0.003 \\
    4--5   & -1.529 &  4.688 &  5.340 &  1.018 &  0.346 &  0.341 &  0.125 &  0.014 &  0.132 &  0.149 &  0.095 &  0.050 &  0.033 &  0.015 &  0.016 \\
    5--6   & -0.807 &  1.180 &  1.018 &  0.617 &  0.139 &  0.154 &  0.078 &  0.035 &  0.051 &  0.050 &  0.031 &  0.017 &  0.011 &  0.006 &  0.007 \\
    6--7   & -0.005 &  0.613 &  0.346 &  0.139 &  0.222 &  0.112 &  0.080 &  0.047 &  0.033 &  0.028 &  0.015 &  0.008 &  0.007 &  0.005 &  0.004 \\
    7--9   &  0.345 &  0.753 &  0.341 &  0.154 &  0.112 &  0.345 &  0.117 &  0.071 &  0.043 &  0.034 &  0.017 &  0.010 &  0.009 &  0.006 &  0.006 \\
    9--12  &  0.632 &  0.555 &  0.125 &  0.078 &  0.080 &  0.117 &  0.263 &  0.073 &  0.038 &  0.029 &  0.013 &  0.007 &  0.008 &  0.006 &  0.006 \\
    12--15 &  0.507 &  0.317 &  0.014 &  0.035 &  0.047 &  0.071 &  0.073 &  0.111 &  0.023 &  0.015 &  0.007 &  0.004 &  0.005 &  0.004 &  0.003 \\
    15--18 & -0.124 &  0.158 &  0.132 &  0.051 &  0.033 &  0.043 &  0.038 &  0.023 &  0.038 &  0.016 &  0.008 &  0.005 &  0.004 &  0.003 &  0.003 \\
    18--22 & -0.309 &  0.084 &  0.149 &  0.050 &  0.028 &  0.034 &  0.029 &  0.015 &  0.016 &  0.029 &  0.009 &  0.005 &  0.004 &  0.003 &  0.003 \\
    22--26 & -0.213 &  0.047 &  0.095 &  0.031 &  0.015 &  0.017 &  0.013 &  0.007 &  0.008 &  0.009 &  0.009 &  0.003 &  0.002 &  0.001 &  0.001 \\
    26--30 & -0.109 &  0.027 &  0.050 &  0.017 &  0.008 &  0.010 &  0.007 &  0.004 &  0.005 &  0.005 &  0.003 &  0.003 &  0.001 &  0.001 &  0.001 \\
    30--36 & -0.045 &  0.030 &  0.033 &  0.011 &  0.007 &  0.009 &  0.008 &  0.005 &  0.004 &  0.004 &  0.002 &  0.001 &  0.003 &  0.001 &  0.001 \\
    36--42 & -0.032 &  0.011 &  0.015 &  0.006 &  0.005 &  0.006 &  0.006 &  0.004 &  0.003 &  0.003 &  0.001 &  0.001 &  0.001 &  0.001 &  0.000 \\
    42--50 & -0.052 &  0.003 &  0.016 &  0.007 &  0.004 &  0.006 &  0.006 &  0.003 &  0.003 &  0.003 &  0.001 &  0.001 &  0.001 &  0.000 &  0.001 \\
    \hline
    \end{tabular} }
\caption{Covariance matrix for the extracted neutrino flux in the forward horn current (FHC) beam. The bin boundaries are in units of GeV. The covariance elements are in units of \numu~/ m$^2$ / 1e6 POT.}
\label{cov:FHC-nu-flux-NOMAD}
\end{table}

\begin{table}[tb]
    {\renewcommand{\tabcolsep}{0.1cm} \begin{tabular}{c|ccccccccccccccc}
    \hline
           &   2--3 &   3--4 &   4--5 &   5--6 &   6--7 &   7--9 &  9--12 & 12--15 & 15--18 & 18--22 & 22--26 & 26--30 & 30--36 & 36--42 & 42--50 \\
    \hline
    2--3   & 29.782 & 10.906 &  6.980 &  3.919 &  3.028 &  3.502 &  0.362 & -0.347 & -0.445 & -0.333 & -0.258 & -0.140 & -0.072 & -0.042 & -0.038 \\
    3--4   & 10.906 & 24.339 &  5.431 &  3.160 &  2.597 &  3.060 &  0.624 & -0.089 & -0.275 & -0.213 & -0.184 & -0.105 & -0.051 & -0.030 & -0.029 \\
    4--5   &  6.980 &  5.431 & 16.906 &  2.251 &  1.728 &  2.149 &  0.524 & -0.085 & -0.187 & -0.139 & -0.162 & -0.087 & -0.044 & -0.026 & -0.026 \\
    5--6   &  3.919 &  3.160 &  2.251 & 11.440 &  1.079 &  1.464 &  0.437 & -0.037 & -0.149 & -0.119 & -0.125 & -0.069 & -0.037 & -0.021 & -0.021 \\
    6--7   &  3.028 &  2.597 &  1.728 &  1.079 &  8.845 &  1.327 &  0.399 & -0.053 & -0.140 & -0.108 & -0.127 & -0.068 & -0.035 & -0.018 & -0.020 \\
    7--9   &  3.502 &  3.060 &  2.149 &  1.464 &  1.327 & 23.211 &  0.789 &  0.186 & -0.020 & -0.034 & -0.082 & -0.048 & -0.021 & -0.010 & -0.016 \\
    9--12  &  0.362 &  0.624 &  0.524 &  0.437 &  0.399 &  0.789 & 19.332 &  0.530 &  0.293 &  0.201 &  0.080 &  0.042 &  0.034 &  0.019 &  0.011 \\
    12--15 & -0.347 & -0.089 & -0.085 & -0.037 & -0.053 &  0.186 &  0.530 &  6.570 &  0.300 &  0.204 &  0.140 &  0.068 &  0.046 &  0.023 &  0.019 \\
    15--18 & -0.445 & -0.275 & -0.187 & -0.149 & -0.140 & -0.020 &  0.293 &  0.300 &  2.010 &  0.151 &  0.101 &  0.051 &  0.033 &  0.017 &  0.015 \\
    18--22 & -0.333 & -0.213 & -0.139 & -0.119 & -0.108 & -0.034 &  0.201 &  0.204 &  0.151 &  0.997 &  0.073 &  0.037 &  0.025 &  0.013 &  0.011 \\
    22--26 & -0.258 & -0.184 & -0.162 & -0.125 & -0.127 & -0.082 &  0.080 &  0.140 &  0.101 &  0.073 &  0.270 &  0.031 &  0.019 &  0.010 &  0.009 \\
    26--30 & -0.140 & -0.105 & -0.087 & -0.069 & -0.068 & -0.048 &  0.042 &  0.068 &  0.051 &  0.037 &  0.031 &  0.072 &  0.010 &  0.005 &  0.005 \\
    30--36 & -0.072 & -0.051 & -0.044 & -0.037 & -0.035 & -0.021 &  0.034 &  0.046 &  0.033 &  0.025 &  0.019 &  0.010 &  0.046 &  0.003 &  0.003 \\
    36--42 & -0.042 & -0.030 & -0.026 & -0.021 & -0.018 & -0.010 &  0.019 &  0.023 &  0.017 &  0.013 &  0.010 &  0.005 &  0.003 &  0.018 &  0.002 \\
    42--50 & -0.038 & -0.029 & -0.026 & -0.021 & -0.020 & -0.016 &  0.011 &  0.019 &  0.015 &  0.011 &  0.009 &  0.005 &  0.003 &  0.002 &  0.011 \\
    \hline
    \end{tabular} }
\caption{Covariance matrix for the extracted antineutrino flux in the forward horn current (FHC) beam. The bin boundaries are in units of GeV. The covariance elements
are in units of \numubar~/ m$^2$ / 1e6 POT and scaled by a factor of $10^2$.}
\label{cov:FHC-anti-flux}
\end{table}

\begin{table}[tb]
    {\renewcommand{\tabcolsep}{0.1cm} \begin{tabular}{c|ccccccccccccccc}
    \hline
           &    2--3 &   3--4 &   4--5 &   5--6 &   6--7 &   7--9 &  9--12 & 12--15 & 15--18 & 18--22 & 22--26 & 26--30 & 30--36 & 36--42 & 42--50 \\
    \hline
    2--3   & 136.690 & 41.023 &  7.723 &  0.381 &  0.308 &  0.078 &  0.216 &  0.081 & -0.049 & -0.137 & -0.089 & -0.049 & -0.013 & -0.022 & -0.018 \\
    3--4   &  41.023 & 81.064 &  9.207 &  1.256 &  0.603 &  0.404 &  0.201 &  0.113 &  0.067 &  0.019 & -0.001 & -0.004 &  0.002 &  0.001 & -0.002 \\
    4--5   &   7.723 &  9.207 & 11.534 &  0.802 &  0.330 &  0.245 &  0.061 &  0.048 &  0.060 &  0.054 &  0.025 &  0.012 &  0.005 &  0.007 &  0.005 \\
    5--6   &   0.381 &  1.256 &  0.802 &  0.962 &  0.079 &  0.071 &  0.021 &  0.013 &  0.013 &  0.011 &  0.006 &  0.003 &  0.000 &  0.001 &  0.001 \\
    6--7   &   0.308 &  0.603 &  0.330 &  0.079 &  0.370 &  0.033 &  0.011 &  0.007 &  0.006 &  0.004 &  0.002 &  0.001 &  0.000 &  0.000 &  0.000 \\
    7--9   &   0.078 &  0.404 &  0.245 &  0.071 &  0.033 &  0.521 &  0.014 &  0.008 &  0.006 &  0.004 &  0.002 &  0.001 &  0.000 &  0.000 &  0.000 \\
    9--12  &   0.216 &  0.201 &  0.061 &  0.021 &  0.011 &  0.014 &  0.338 &  0.005 &  0.003 &  0.002 &  0.001 &  0.000 &  0.000 &  0.000 &  0.000 \\
    12--15 &   0.081 &  0.113 &  0.048 &  0.013 &  0.007 &  0.008 &  0.005 &  0.091 &  0.002 &  0.001 &  0.001 &  0.000 &  0.000 &  0.000 &  0.000 \\
    15--18 &  -0.049 &  0.067 &  0.060 &  0.013 &  0.006 &  0.006 &  0.003 &  0.002 &  0.029 &  0.002 &  0.001 &  0.000 &  0.000 &  0.000 &  0.000 \\
    18--22 &  -0.137 &  0.019 &  0.054 &  0.011 &  0.004 &  0.004 &  0.002 &  0.001 &  0.002 &  0.014 &  0.001 &  0.000 &  0.000 &  0.000 &  0.000 \\
    22--26 &  -0.089 & -0.001 &  0.025 &  0.006 &  0.002 &  0.002 &  0.001 &  0.001 &  0.001 &  0.001 &  0.004 &  0.000 &  0.000 &  0.000 &  0.000 \\
    26--30 &  -0.049 & -0.004 &  0.012 &  0.003 &  0.001 &  0.001 &  0.000 &  0.000 &  0.000 &  0.000 &  0.000 &  0.001 &  0.000 &  0.000 &  0.000 \\
    30--36 &  -0.013 &  0.002 &  0.005 &  0.000 &  0.000 &  0.000 &  0.000 &  0.000 &  0.000 &  0.000 &  0.000 &  0.000 &  0.001 &  0.000 &  0.000 \\
    36--42 &  -0.022 &  0.001 &  0.007 &  0.001 &  0.000 &  0.000 &  0.000 &  0.000 &  0.000 &  0.000 &  0.000 &  0.000 &  0.000 &  0.001 &  0.000 \\
    42--50 &  -0.018 & -0.002 &  0.005 &  0.001 &  0.000 &  0.000 &  0.000 &  0.000 &  0.000 &  0.000 &  0.000 &  0.000 &  0.000 &  0.000 &  0.000 \\
    \hline
    \end{tabular} }
\caption{Covariance matrix for the extracted antineutrino flux in the reverse horn current (RHC) beam. The bin boundaries are in units of GeV. The covariance elements are in units of \numubar~/ m$^2$ / 1e6 POT.}
\label{cov:RHC-anti-flux}
\end{table}

\begin{table}[tb]
    {\renewcommand{\tabcolsep}{0.1cm} \begin{tabular}{c|ccccccccccccccc}
    \hline
           &   2--3 &   3--4 &   4--5 &   5--6 &   6--7 &   7--9 &  9--12 & 12--15 & 15--18 & 18--22 & 22--26 & 26--30 & 30--36 & 36--42 & 42--50 \\
    \hline
    2--3   & 25.635 &  6.856 &  5.532 &  5.118 &  5.019 &  7.871 &  5.803 &  1.770 &  0.347 & -0.067 & -0.068 & -0.193 & -0.187 &  0.163 &  0.187 \\
    3--4   &  6.856 & 25.219 &  5.769 &  5.002 &  5.388 &  8.801 &  7.363 &  3.075 &  1.281 &  0.777 &  0.344 &  0.063 &  0.043 &  0.269 &  0.316 \\
    4--5   &  5.532 &  5.769 & 20.512 &  4.494 &  4.617 &  8.025 &  6.941 &  3.082 &  1.407 &  0.950 &  0.439 &  0.114 &  0.108 &  0.270 &  0.318 \\
    5--6   &  5.118 &  5.002 &  4.494 & 13.886 &  4.206 &  7.380 &  6.273 &  2.479 &  1.015 &  0.653 &  0.252 &  0.011 &  0.028 &  0.217 &  0.254 \\
    6--7   &  5.019 &  5.388 &  4.617 &  4.206 & 15.103 &  8.491 &  7.322 &  2.995 &  1.328 &  1.000 &  0.410 &  0.043 &  0.078 &  0.275 &  0.327 \\
    7--9   &  7.871 &  8.801 &  8.025 &  7.380 &  8.491 & 45.734 & 14.318 &  6.868 &  3.725 &  2.767 &  1.342 &  0.501 &  0.462 &  0.609 &  0.741 \\
    9--12  &  5.803 &  7.363 &  6.941 &  6.273 &  7.322 & 14.318 & 49.374 &  8.873 &  5.286 &  4.155 &  2.097 &  0.973 &  0.835 &  0.753 &  0.946 \\
    12--15 &  1.770 &  3.075 &  3.082 &  2.479 &  2.995 &  6.868 &  8.873 & 18.746 &  3.921 &  3.160 &  1.637 &  0.859 &  0.704 &  0.492 &  0.642 \\
    15--18 &  0.347 &  1.281 &  1.407 &  1.015 &  1.328 &  3.725 &  5.286 &  3.921 &  7.821 &  2.222 &  1.172 &  0.643 &  0.528 &  0.319 &  0.421 \\
    18--22 & -0.067 &  0.777 &  0.950 &  0.653 &  1.000 &  2.767 &  4.155 &  3.160 &  2.222 &  4.877 &  0.951 &  0.503 &  0.424 &  0.259 &  0.347 \\
    22--26 & -0.068 &  0.344 &  0.439 &  0.252 &  0.410 &  1.342 &  2.097 &  1.637 &  1.172 &  0.951 &  1.450 &  0.287 &  0.234 &  0.136 &  0.174 \\
    26--30 & -0.193 &  0.063 &  0.114 &  0.011 &  0.043 &  0.501 &  0.973 &  0.859 &  0.643 &  0.503 &  0.287 &  0.515 &  0.146 &  0.070 &  0.089 \\
    30--36 & -0.187 &  0.043 &  0.108 &  0.028 &  0.078 &  0.462 &  0.835 &  0.704 &  0.528 &  0.424 &  0.234 &  0.146 &  0.436 &  0.058 &  0.072 \\
    36--42 &  0.163 &  0.269 &  0.270 &  0.217 &  0.275 &  0.609 &  0.753 &  0.492 &  0.319 &  0.259 &  0.136 &  0.070 &  0.058 &  0.223 &  0.055 \\
    42--50 &  0.187 &  0.316 &  0.318 &  0.254 &  0.327 &  0.741 &  0.946 &  0.642 &  0.421 &  0.347 &  0.174 &  0.089 &  0.072 &  0.055 &  0.302 \\
\hline
\end{tabular} }
\caption{Covariance matrix for the extracted neutrino flux in the reverse horn current (RHC) beam. The bin boundaries are in units of GeV. The covariance elements are in units of \numu~/ m$^2$ / 1e6 POT and scaled by a factor of $10^2$.}
\label{cov:RHC-nu-flux}
\end{table}

\begin{table}[tb]
    {\renewcommand{\tabcolsep}{0.1cm} \begin{tabular}{c|ccccccccccccccc}
    \hline
           &   2--3 &   3--4 &   4--5 &   5--6 &   6--7 &   7--9 &  9--12 & 12--15 & 15--18 & 18--22 & 22--26 & 26--30 & 30--36 & 36--42 & 42--50 \\
    \hline
    2--3   & 12.016 &  4.494 &  0.063 & -0.621 & -0.155 &  0.284 & -0.001 & -0.668 & -1.215 & -1.407 & -1.891 & -2.439 & -1.595 & -0.932 & -0.255 \\
    3--4   &  4.494 &  3.598 &  1.038 &  0.263 &  0.218 &  0.257 &  0.010 & -0.325 & -0.518 & -0.658 & -0.864 & -1.113 & -0.794 & -0.567 & -0.413 \\
    4--5   &  0.063 &  1.038 &  2.376 &  0.723 &  0.407 &  0.227 &  0.012 & -0.072 &  0.162 &  0.193 &  0.299 &  0.279 &  0.097 & -0.050 & -0.152 \\
    5--6   & -0.621 &  0.263 &  0.723 &  1.363 &  0.302 &  0.174 &  0.009 & -0.026 &  0.121 &  0.097 &  0.145 &  0.128 & -0.020 & -0.074 & -0.150 \\
    6--7   & -0.155 &  0.218 &  0.407 &  0.302 &  1.094 &  0.147 &  0.006 & -0.039 &  0.027 & -0.022 & -0.017 & -0.065 & -0.127 & -0.114 & -0.155 \\
    7--9   &  0.284 &  0.257 &  0.227 &  0.174 &  0.147 &  1.053 &  0.005 & -0.076 & -0.038 & -0.095 & -0.139 & -0.199 & -0.210 & -0.153 & -0.115 \\
    9--12  & -0.001 &  0.010 &  0.012 &  0.009 &  0.006 &  0.005 &  0.728 &  0.004 &  0.003 &  0.002 &  0.003 &  0.006 &  0.002 &  0.005 &  0.005 \\
    12--15 & -0.668 & -0.325 & -0.072 & -0.026 & -0.039 & -0.076 &  0.004 &  0.883 &  0.127 &  0.160 &  0.221 &  0.319 &  0.251 &  0.180 &  0.103 \\
    15--18 & -1.215 & -0.518 &  0.162 &  0.121 &  0.027 & -0.038 &  0.003 &  0.127 &  1.245 &  0.523 &  0.705 &  0.842 &  0.585 &  0.381 &  0.317 \\
    18--22 & -1.407 & -0.658 &  0.193 &  0.097 & -0.022 & -0.095 &  0.002 &  0.160 &  0.523 &  1.778 &  1.088 &  1.275 &  0.917 &  0.608 &  0.555 \\
    22--26 & -1.891 & -0.864 &  0.299 &  0.145 & -0.017 & -0.139 &  0.003 &  0.221 &  0.705 &  1.088 &  2.848 &  1.784 &  1.277 &  0.839 &  0.733 \\
    26--30 & -2.439 & -1.113 &  0.279 &  0.128 & -0.065 & -0.199 &  0.006 &  0.319 &  0.842 &  1.275 &  1.784 &  3.783 &  1.568 &  1.035 &  0.875 \\
    30--36 & -1.595 & -0.794 &  0.097 & -0.020 & -0.127 & -0.210 &  0.002 &  0.251 &  0.585 &  0.917 &  1.277 &  1.568 &  2.650 &  0.790 &  0.679 \\
    36--42 & -0.932 & -0.567 & -0.050 & -0.074 & -0.114 & -0.153 &  0.005 &  0.180 &  0.381 &  0.608 &  0.839 &  1.035 &  0.790 &  2.281 &  0.516 \\
    42--50 & -0.255 & -0.413 & -0.152 & -0.150 & -0.155 & -0.115 &  0.005 &  0.103 &  0.317 &  0.555 &  0.733 &  0.875 &  0.679 &  0.516 &  3.133 \\
    \hline
    \end{tabular} }
\caption{Covariance matrix for the isoscalar corrected extracted cross section for neutrinos. The bin boundaries are in units of GeV. The covariance elements are in units of $10^{-38}$ cm$^2$ / GeV / nucleon and scaled by a factor of $10^3$.}
\label{cov:FHC-nu-xsec-NOMAD}
\end{table}

\begin{table}[tb]
    {\renewcommand{\tabcolsep}{0.1cm} \begin{tabular}{c|ccccccccccccccc}
    \hline
           &   2--3 &   3--4 &   4--5 &   5--6 &   6--7 &   7--9 &  9--12 & 12--15 & 15--18 & 18--22 & 22--26 & 26--30 & 30--36 & 36--42 & 42--50 \\
    \hline
    2--3   &  3.144 &  0.699 & -0.031 &  0.009 &  0.148 &  0.150 &  0.005 & -0.134 & -0.294 & -0.434 & -0.549 & -0.627 & -0.652 & -0.580 & -0.802 \\
    3--4   &  0.699 &  1.686 &  0.157 &  0.051 &  0.093 &  0.064 & -0.010 & -0.050 & -0.100 & -0.166 & -0.233 & -0.275 & -0.240 & -0.211 & -0.324 \\
    4--5   & -0.031 &  0.157 &  1.626 &  0.085 &  0.053 &  0.006 & -0.019 &  0.016 &  0.065 &  0.086 &  0.080 &  0.062 &  0.092 &  0.133 &  0.124 \\
    5--6   &  0.009 &  0.051 &  0.085 &  1.492 &  0.034 &  0.021 & -0.006 &  0.009 &  0.018 &  0.021 &  0.016 & -0.020 & -0.001 &  0.023 & -0.021 \\
    6--7   &  0.148 &  0.093 &  0.053 &  0.034 &  1.397 &  0.037 & -0.003 & -0.008 & -0.018 & -0.026 & -0.043 & -0.088 & -0.090 & -0.034 & -0.100 \\
    7--9   &  0.150 &  0.064 &  0.006 &  0.021 &  0.037 &  1.488 &  0.002 & -0.007 & -0.020 & -0.033 & -0.044 & -0.074 & -0.075 & -0.042 & -0.105 \\
    9--12  &  0.005 & -0.010 & -0.019 & -0.006 & -0.003 &  0.002 &  1.185 & -0.000 & -0.002 & -0.001 &  0.001 &  0.003 & -0.002 & -0.003 & -0.001 \\
    12--15 & -0.134 & -0.050 &  0.016 &  0.009 & -0.008 & -0.007 & -0.000 &  1.451 &  0.037 &  0.051 &  0.054 &  0.060 &  0.085 &  0.072 &  0.067 \\
    15--18 & -0.294 & -0.100 &  0.065 &  0.018 & -0.018 & -0.020 & -0.002 &  0.037 &  1.456 &  0.124 &  0.153 &  0.168 &  0.169 &  0.160 &  0.206 \\
    18--22 & -0.434 & -0.166 &  0.086 &  0.021 & -0.026 & -0.033 & -0.001 &  0.051 &  0.124 &  1.823 &  0.281 &  0.288 &  0.237 &  0.277 &  0.362 \\
    22--26 & -0.549 & -0.233 &  0.080 &  0.016 & -0.043 & -0.044 &  0.001 &  0.054 &  0.153 &  0.281 &  2.776 &  0.398 &  0.297 &  0.345 &  0.483 \\
    26--30 & -0.627 & -0.275 &  0.062 & -0.020 & -0.088 & -0.074 &  0.003 &  0.060 &  0.168 &  0.288 &  0.398 &  4.293 &  0.423 &  0.397 &  0.598 \\
    30--36 & -0.652 & -0.240 &  0.092 & -0.001 & -0.090 & -0.075 & -0.002 &  0.085 &  0.169 &  0.237 &  0.297 &  0.423 &  2.627 &  0.363 &  0.485 \\
    36--42 & -0.580 & -0.211 &  0.133 &  0.023 & -0.034 & -0.042 & -0.003 &  0.072 &  0.160 &  0.277 &  0.345 &  0.397 &  0.363 &  3.571 &  0.548 \\
    42--50 & -0.802 & -0.324 &  0.124 & -0.021 & -0.100 & -0.105 & -0.001 &  0.067 &  0.206 &  0.362 &  0.483 &  0.598 &  0.485 &  0.548 &  6.478 \\
    \hline
    \end{tabular} }
\caption{Covariance matrix for the isoscalar corrected extracted cross section for antineutrinos. The bin boundaries are in units of GeV. The covariance elements are in units of $10^{-38}$ cm$^2$ / GeV / nucleon and scaled by a factor of $10^3$.}
\label{cov:RHC-anti-xsec-isoscalar}
\end{table}